\documentclass[3p,12pt]{elsarticle}
\usepackage{tabu}
\usepackage[most]{tcolorbox}
\usepackage{float}
\usepackage{graphicx}
\usepackage{subfigure} 
\usepackage{amssymb}
\usepackage[utf8]{inputenc}
\usepackage{pdflscape}

\usepackage{graphicx}
\usepackage{amssymb}
\usepackage[utf8]{inputenc}
\usepackage{bm}
\usepackage{lineno}
\usepackage{hyperref}
\usepackage[capitalize,compress]{cleveref}  

\usepackage{xcolor}
\usepackage{multirow}
\usepackage{multicol}
\usepackage{booktabs}

\newcommand*\patchAmsMathEnvironmentForLineno[1]{%
	\expandafter\let\csname old#1\expandafter\endcsname\csname #1\endcsname
	\expandafter\let\csname oldend#1\expandafter\endcsname\csname end#1\endcsname
	\renewenvironment{#1}%
	{\linenomath\csname old#1\endcsname}%
	{\csname oldend#1\endcsname\endlinenomath}}%
\newcommand*\patchBothAmsMathEnvironmentsForLineno[1]{%
	\patchAmsMathEnvironmentForLineno{#1}%
	\patchAmsMathEnvironmentForLineno{#1*}}%
\AtBeginDocument{%
	\patchBothAmsMathEnvironmentsForLineno{equation}%
	\patchBothAmsMathEnvironmentsForLineno{align}%
	\patchBothAmsMathEnvironmentsForLineno{flalign}%
	\patchBothAmsMathEnvironmentsForLineno{alignat}%
	\patchBothAmsMathEnvironmentsForLineno{gather}%
	\patchBothAmsMathEnvironmentsForLineno{multline}%
}

\begin{document}
\begin{frontmatter}
\title{A thermodynamic framework for unified continuum models for the healing of damaged soft biological tissue}
		

		
		\author[label1]{Di Zuo}
		\author[label1]{Yiqian He\corref{cor1}}
		\author[label3]{Stéphane Avril}
		\author[label1]{Haitian Yang}
	    \author[label2]{Klaus Hackl}
		\address[label1]{State Key Lab of Structural Analysis for Industrial equipment, Department of Engineering Mechanics, Dalian University of Technology, Dalian 116024, P.R. China }
		\address[label2]{Mechanik–Materialtheorie, Ruhr-Universität Bochum, Bochum, Germany}
		\address[label3]{Mines Saint-Etienne, University of Lyon, University Jean Monnet, Inserm, Sainbiose U1059, F-42023 Saint-Etienne, France}
		\cortext[cor1]{Corresponding author:heyiqian@dlut.edu.cn}

\begin{abstract}
When they are damaged or injured, soft biological tissues are able to self-repair and heal.
Mechanics is critical during the healing process, as the damaged extracellular matrix (ECM) tends to be replaced with a new undamaged ECM supporting homeostatic stresses.
Computational modeling has been commonly used to simulate the healing process. However, there
is a pressing need to have a unified thermodynamics theory for healing. From the viewpoint
of continuum damage mechanics, some key parameters related to healing processes, for instance,
the volume fraction of newly grown soft tissue and the growth deformation, can be
regarded as internal variables and have related evolution equations. This paper is aiming
to establish this unified framework inspired by thermodynamics for continuum damage
models for the healing of soft biological tissues. The significant advantage of the proposed model
is that no \textit{ad hoc} equations are required for describing the healing process. Therefore, this new model is more concise and offers a universal approach to simulate the healing process.
Three numerical examples are provided to demonstrate the effectiveness of the proposed model, which is in good agreement with the existing works, including an application for balloon angioplasty in an arteriosclerotic artery with a fiber cap.
\end{abstract}
		
\end{frontmatter}
	
\section{Introduction}
\label{S:1}
Soft biological tissue, such as arteries, skin, ligaments, and tendons, has the ability to grow and change through the formation of new constituents and the removal of old constituents \cite{rao2011modeling}. Understanding the underlying mechanisms of the healing of damaged soft tissue has important applications, for instance, the accurate prediction of the rupture risk of an aortic aneurysm is critical to improve clinical treatment planning \cite{Gasser2017}, the understanding of short-term and long-term damage evolution in the interaction with medical devices for soft tissue is essential for the optimization of these devices \cite{Gasser2017}, and the modeling of wound healing in the skin can improve wound and scar treatment \cite{valero2015modeling,comellas2016homeostatic}.

The healing of soft biological tissue is a complex biochemical and biomechanical process of self-recovering or self-repairing the injured or damaged extracellular matrix (ECM), and is usually divided into four stages: haemostasis, inflammation, proliferation, and remodeling. These four stages were described in great detail by Comellas et al. \cite{comellas2016homeostatic} and Cumming et al. \cite{cumming2009mathematical}. It was reported that the first three stages (from haemostasis to proliferation) may last several weeks and that the final stage of remodeling may last from weeks to years. This last stage consists of continuous turnover (synthesis and degradation) of the ECM simultaneously with the production of scar tissue. The mechanical loading is to be proved to have a significant impact on the speed and efficiency of healing, although the underlying detailed mechanobiological mechanisms involved are not fully clear \cite{comellas2016homeostatic}.

Computational modeling can provide insight into the healing of soft tissues from both short-term and long-term perspectives, and has become more popular for intense research, since experimental research is very time consuming and always involves ethical issues arising from the use of living samples. Generally, there are two types of approaches \cite{valero2015modeling,buganza2016computational}: The first type focuses on the underlying cellular and biochemical mechanisms based on continuum or hybrid discrete/continuum approaches, including the simulation of wound contraction \cite{buganza2016computational,javierre2009numerical} and angiogenesis \cite{schugart2008wound}, providing means to reveal the underlying mechanism, usually from a microscopic view. The other type, more phenomenological, focuses on the change in the material properties of tissue during the remodeling phase. For instance, Comellas et al. developed a homeostasis-driven turnover remodeling model for healing in soft tissues based on continuum damage mechanics \cite{comellas2016homeostatic}. Moreover, some studies focus on modeling the specified remodeling process, e.g., the collagen fiber reorientation \cite{kuhl2005remodeling}, {continuous turnover of constituents} \cite{kuhl2005remodeling} and constrained mixture computational method \cite{humphrey2002constrained,latorre2018critical,dimitrijevic2008method}. 

Despite the existing works introduced above, the computational modeling of healing is still challenging, a main drawback of the current healing models being that some \textit{ad hoc} equations, based on different assumptions, have to be employed to describe the change in variables, resulting in a significant increase in modeling complexity. 
Continuum damage mechanics (CDM), which is consistent with an open-system thermodynamics framework, provides a powerful approach to capture the continuous turnover of tissue. In our previous work \cite{he2019gradient,zuo2020threedimensional}, a nonlocal continuum healing model was presented by combining a gradient-enhanced damage model and a temporally homogenized G\&R model.
Instead of introducing a mechanobiological model to describe the G\&R process, e.g., the temporally homogenized {growth and remodeling (G\&R)} model, a more general and unified approach is newly presented in this paper. 
The core idea is that, from the viewpoint of CDM, the parameters related to the healing process can be regarded as internal variables, in the same way as the damage variable. 
Therefore, it is possible to establish a more rigorous and concise unified model without any \textit{ad hoc} equations, including the evolution equations for the growth and remodeling based on strict thermodynamic considerations.
To the best of the authors' knowledge, there appears to be no work related to the model proposed in this paper to date.

Based on the above considerations, a new unified continuum damage model is first established in this paper. The proposed theoretical framework provides a more convenient computational model without any \textit{ad hoc} equations. 
A numerical simulation based on the newly established damage models will result in a powerful tool for predicting and understanding the mechanism of the self-healing behavior in soft biological tissues, particularly in regard to diseases or wounds such as aneurysms or skin wounds, and will help to improve related treatment methods.  

The paper is organized as follows:
\cref{S:2} introduces the framework of the unified damage model for healing, including the basic kinematics in \cref{S:2.1}, thermodynamic modeling of growth in \cref{S:2.2}, coupling to remodeling in \cref{S:2.3}, coupling to damage in \cref{S:2.4}, summary of the evolution equations of healing in \cref{S:2.51}, gradient-enhanced nonlocal damage model in \cref{S:2.6}, total potential energy and variational form in \cref{S:2.5}, and constitutive model in \cref{S:2.8}. \cref{S:3} provides numerical examples to demonstrate the effectiveness of the proposed model. Finally, discussions and conclusions are given in \cref{S:4}.

\section{The thermodynamic framework of the unified damage model for healing}
\label{S:2}
\subsection{Basic kinematics}
\label{S:2.1}
Let $\bm x=\bm \varphi(\bm X,t)$ describe the motion of the. This equation transforms referential placements $\bm x\in \kappa(0)$ into their spatial counterparts $\bm x\in \kappa(t)$, where $\kappa(0)$ and $\kappa(t)$ are the initial reference configuration and current configuration, respectively. The deformation gradient and the Jacobian, which maps the referential volume $dV$ onto the current volume $dv$, are defined as

\begin{equation}
\label{equ:1}
{\bm {F}} = {\nabla _{\bm{X}}}{\bm{\varphi }},
\end{equation}

\begin{equation}
\label{equ.2}
J=\frac{dv}{dV}=\det(\bm F).
\end{equation}

We introduce a variational approach for the description of inelastic processes that rests on thermodynamic extremal principles. For this purpose, let us consider a physical system described by (sets of) external, i.e., controllable, state variables, in our case given by the deformation gradient $\bm{F}$, and internal state variables $\bm{z}$.

We assume that the system behavior may be defined using only two scalar potentials: free energy $\psi(\bm{F},\bm{z})$ and dissipation potential $\Delta(\bm{z},\dot{\bm{z}})$.
The deformation $\bm{x}$ is given by the minimization of energy as
\begin{equation}
\label{eq2}
\inf_{\bm{x}}\left\{{\int_\Omega \psi(\bm{F},\bm{z}) {\rm d} V + f_\mathrm{ext}(\bm{x})} | {\bm{x}=\bm{x}_0 \;\mbox{on}\; \partial\Omega}\right\},
\end{equation}
where $f_\mathrm{ext}(\bm{x})$ denotes the potential of external driving forces. The evolution of the internal variables is described by the Biot equation.
\begin{equation}
\label{eq3}
\frac{\partial \psi}{\partial {\bm{z}}}+\frac{\partial \Delta}{\partial \dot{\bm{z}}} = {\bf 0}.
\end{equation}

Note that \cref{eq3} may be written as a stationarity condition of the minimization problem
\begin{equation}
\label{eq4}
\inf_{\dot{\bm{z}}}\left\{{\dot{\Psi} + \Delta}\right\}.
\end{equation}

\subsection{Thermodynamic modeling of growth}
\label{S:2.2}

Let us assume that the local state of tissue growth is given by an internal variable of an inelastic deformation gradient $\bm{z}=\bm{F}_\mathrm{g}$, such that the total deformation gradient is given as
\begin{equation}
\label{eq5}
\bm{F}=\bm{F}_\mathrm{e} \cdot \bm{F}_\mathrm{g},
\end{equation}
where $\bm{F}_\mathrm{e}$ denotes the part of the deformation gradient given by elastic straining. The elastic free energy depends on $\bm{F}_\mathrm{e}$ only. Hence we have
\begin{equation}
\label{eq6}
\psi_\mathrm{el}(\bm{F}_\mathrm{e}) = \psi_\mathrm{el}(\bm{F} \cdot \bm{F}_\mathrm{g}^{-1}).
\end{equation}

In a material that has already undergone growth, the free energy must be related to the volume of the grown tissue, i.e., premultiplied by $J_\mathrm{g} = \det (\bm{F}_\mathrm{g}$). Moreover, we introduce a constant term $\Delta\psi_\mathrm{ph}$ contained in the free energy, which we denote as the physiological potential. We assume that $\Delta\psi_\mathrm{ph}$ can be influenced by physiological processes to stimulate growth where necessary. 
With this notion, the free energy takes the form
\begin{equation}
\label{eq7}
\psi(\bm{F},\bm{F}_\mathrm{g}) = J_\mathrm{g} \, \left( \psi_\mathrm{el}(\bm{F} \cdot \bm{F}_\mathrm{g}^{-1}) + \Delta\psi_\mathrm{ph} \right).
\end{equation}

We obtain the first Piola-Kirchhoff stress tensor as
\begin{equation}
\label{eq8}
\bm{P} = \frac{\partial \psi}{\partial \bm{F}} = J_\mathrm{g} \, \frac{\partial \psi_\mathrm{el}}{\partial \bm{F}_\mathrm{e}} \cdot \bm{F}_\mathrm{g}^{-\mathrm{T}},
\end{equation}
and the Cauchy stress tensor as
\begin{equation}
\label{eq9}
\bm\sigma = \frac{1}{J} \, \bm{P} \cdot \bm{F}^\mathrm{T} = \frac{1}{J_\mathrm{e}} \, \frac{\partial \psi_\mathrm{el}}{\partial \bm{F}_\mathrm{e}} \cdot \bm{F}_\mathrm{e}^\mathrm{T},
\end{equation}
where $J = \det (\bm{F}$) and $J_\mathrm{e} = \det (\bm{F}_\mathrm{e}$). Material frame indifference requires that $\psi_\mathrm{el}$ factors through the right Cauchy-Green tensor $\bm{C}_\mathrm{e} = \bm{F}_\mathrm{e}^\mathrm{T} \cdot \bm{F}_\mathrm{e}$. Employing the relation
\begin{equation}
\label{eq9a}
\frac{\partial \psi_\mathrm{el}}{\partial \bm{F}_\mathrm{e}} = 2 \, \bm{F}_\mathrm{e} \cdot \frac{\partial \psi_\mathrm{el}}{\partial \bm{C}_\mathrm{e}},
\end{equation}
we obtain
\begin{equation}
\label{eq9b}
\bm\sigma = \frac{1}{J_\mathrm{e}} \, 2 \, \bm{F}_\mathrm{e} \cdot \frac{\partial \psi_\mathrm{el}}{\partial \bm{C}_\mathrm{e}} \cdot \bm{F}_\mathrm{e}^\mathrm{T},
\end{equation}
where $\bm\sigma$ is indeed symmetric. Note that $\bm\sigma$ depends on $\bm{F}_\mathrm{e}$ only.

We introduce a thermodynamic driving force associated with growth by
\begin{equation}
\label{eq10}
\bm{q}_\mathrm{g} = - \frac{\partial \psi}{\partial \bm{F}_\mathrm{g}}.
\end{equation}

Employing \cref{eq8} and the fact, that $\frac{\partial J_\mathrm{g}}{\partial \bm{F}_\mathrm{g}} = J_\mathrm{g} \bm{F}_\mathrm{g}^{-\mathrm{T}}$, we obtain
\begin{equation}
\label{eq11}
\bm{q}_\mathrm{g} = - \bm{F}_\mathrm{g}^{-\mathrm{T}} \cdot \left( \psi \, \bm{I} - \bm{F}^\mathrm{T} \cdot \bm{P} \right).
\end{equation}

Note that $\bm{b} = \psi \, \bm{I} - \bm{F}^\mathrm{T} \cdot \bm{P}$ is the Eshelby stress tensor known to be associated with the configuration change.

To close our model, we still have to introduce a dissipation potential. Because we have to respect the material frame indifference once again, we have to formulate this by employing an objective rate. A straightforward choice is the velocity gradient $\bm{L}_\mathrm{g} = \dot{\bm{F}}_\mathrm{g} \cdot \bm{F}_\mathrm{g}^{-1}$. With this notion, we define the dissipation potential as
\begin{equation}
\label{eq12}
\Delta_\mathrm{g}(\bm{L}_\mathrm{g}) = J_\mathrm{g} \, \left( r_\mathrm{g} \, \| \bm{L}_\mathrm{g} \| + \frac{1}{2 M_\mathrm{g}} \, \| \bm{L}_\mathrm{g} \|^2 \right).
\end{equation} 

Note that, similar to the free energy, the dissipation potential has to be premultiplied by $J_\mathrm{g}$. Because of the non-differentiability of $\Delta$ at $\bm{L}_\mathrm{g}=\bm{0}$, \cref{eq3} becomes a differential inclusion and takes the form
\begin{equation}
\label{eq13}
\bm{q}_\mathrm{g} \in J_\mathrm{g} \, \left( r_\mathrm{g} \, \mathrm{sign} \, \bm{L}_\mathrm{g} + \frac{1}{M_\mathrm{g}} \, \bm{L}_\mathrm{g} \right) \cdot \bm{F}_\mathrm{g}^{-\mathrm{T}},
\end{equation}
which has the solution
\begin{equation}
\label{eq14}
\bm{L}_\mathrm{g} = M_\mathrm{g} \left(\frac{1}{J_\mathrm{g}} \,  \| \bm{q}_\mathrm{g} \cdot \bm{F}_\mathrm{g}^\mathrm{T} \| - r_\mathrm{g} \right) _+ \, \mathrm{sign}\left( \bm{q}_\mathrm{g} \cdot \bm{F}_\mathrm{g}^\mathrm{T} \right) ,
\end{equation}
where $(\cdot)_+$ denotes the positive part of the argument. The set-valued sign function of a tensor $\bm{T}$ is defined as
\begin{equation} \label{eq14a} 
\mathrm{sign} (\bm{T}) = \left\lbrace 
\begin{array}{ll}
\left\lbrace \bm{S}, \, \|\bm{S}\| \leq 1 \right\rbrace & \text{for} \; \bm{T}=\bm{0} \\
\left\lbrace \frac{1}{\|\bm{T}|} \,\bm{T} \right\rbrace & \text{for} \; \bm{T}\not=\bm{0} 
\end{array}
\right. .
\end{equation}

We see that $M_\mathrm{g}$ plays the role of a mobility controlling the velocity of the growth process, and $r_\mathrm{g}$ corresponds to a growth limit that is related to a homeostatic state.

A brief calculation gives
\begin{equation}
\label{eq15}
\bar{\bm{q}}_\mathrm{g} := \frac{1}{J_\mathrm{g}} \, \bm{q}_\mathrm{g} \cdot \bm{F}_\mathrm{g}^\mathrm{T} = - \left( \psi_\mathrm{el} + \Delta\psi_\mathrm{ph} \right) \bm{I} + \bm{F}_\mathrm{e}^\mathrm{T} \cdot \frac{\partial \psi_\mathrm{el}}{\partial \bm{F}_\mathrm{e}},
\end{equation}
and \cref{eq14} becomes
\begin{equation}
\label{eq16}
\bm{L}_\mathrm{g} = M_\mathrm{g} \left( \| \bar{\bm{q}}_\mathrm{g} \| - r_\mathrm{g} \right)_+ \, \mathrm{sign}\left( \bar{\bm{q}}_\mathrm{g} \right) .
\end{equation}

Note that $\bar{\bm{q}}_\mathrm{g}$ depends on $\bm{F}_\mathrm{e}$ only. Using \cref{eq9a,eq9b}, we obtain
\begin{equation}
\label{eq17}
\bar{\bm{q}}_\mathrm{g} = - J_\mathrm{e} \,  \bm{I} + 2 \, \bm{C}_\mathrm{e} \cdot \frac{\partial \psi_\mathrm{el}}{\partial \bm{C}_\mathrm{e}} = - \left( \psi_\mathrm{el} + \Delta\psi_\mathrm{ph} \right) \bm{I} + J_\mathrm{e} \, \bm{F}_\mathrm{e}^\mathrm{T} \cdot \bm \sigma \cdot \bm{F}_\mathrm{e}^{-\mathrm{T}}.
\end{equation}

For deformations large enough in order to hold $\| \bar{\bm{q}}_\mathrm{g} \| > r_\mathrm{g}$, the driving forces tend to converge to the hypersurface given by 
\begin{equation}
\label{eq18}
\| \bar{\bm{q}}_\mathrm{g} \| = r_\mathrm{g},
\end{equation}
defining a yield condition for growth. 
Let us investigate this hypersurface closer. A straightforward calculation gives
\begin{equation}
\label{eq18a}
\| \bar{\bm{q}}_\mathrm{g} \|^2 = 3 \left( \psi_\mathrm{el} + \Delta\psi_\mathrm{ph} \right)^2 - 2 J_\mathrm{e} \, \left( \psi_\mathrm{el} + \Delta\psi_\mathrm{ph} \right) \,  \, \bm  + J_\mathrm{e}^2 \, \| \bm{F}_\mathrm{e}^{-1} \cdot \bm \sigma \cdot \bm{F}_\mathrm{e} \|^2.
\end{equation}

We see that \cref{eq18} involves the Cauchy stress $\bm \sigma$ but in general also the elastic deformation gradient $\bm{F}_\mathrm{e}$. Note that the physiological potential $\Delta\psi_\mathrm{ph}$ shifts the growth to higher volumetric stresses.

\subsubsection{The isotropic case}

In this paper, only the case of isotropic growth is considered. For an isotropic tissue, the growth will be volumetric only, i.e., we have 
\begin{equation}
\label{eq19}
\bm{F}_\mathrm{g} = J_\mathrm{g}^{1/3} \, \bm{I}.
\end{equation}

We assume a split of the free energy into a volumetric and an isochoric part as
\begin{equation}
\label{eq20}
\psi_\mathrm{el}(\bm{F}_\mathrm{e}) = U(J_\mathrm{e}) + \psi_\mathrm{iso}(\bar{\bm{F}}_\mathrm{e}),
\end{equation}
where $\bar{\bm{F}}_\mathrm{e}=J_\mathrm{e}^{-1/3}\, \bm{F}_\mathrm{e}$. Employing \cref{eq19}, we obtain
\begin{equation}
\label{eq21}
\bar{\bm{F}}_\mathrm{e} = \bar{\bm{F}} = J^{-1/3}\, \bm{F}.
\end{equation}

A brief calculation using \cref{eq20} gives
\begin{equation}
\label{eq22}
\bm{P} = J_\mathrm{g} \, \left( J_\mathrm{e} \, U^\prime(J_\mathrm{e}) \, \bm{I} + J^{-1/3} \, {\rm dev} \left( \frac{\partial \psi_\mathrm{iso}}{\partial \bar{\bm{F}}} \cdot \bar{\bm{F}}^\mathrm{T} \right) \right)  \cdot \bm{F}^{-\mathrm{T}},
\end{equation}

\begin{equation}
\label{eq23}
\bm \sigma = U^\prime(J_\mathrm{e}) \, \bm{I} + J_\mathrm{g} \, J^{-4/3} \, {\rm dev}  \left( \frac{\partial \psi_\mathrm{iso}}{\partial \bar{\bm{F}}} \cdot \bar{\bm{F}}^\mathrm{T} \right) ,
\end{equation}

\begin{equation}
\label{eq24}
\bar{\bm{q}}_\mathrm{g} = - \left( \psi_\mathrm{el} + \Delta\psi_\mathrm{ph} + J_\mathrm{e} \, U^\prime(J_\mathrm{e}) \right) \bm{I} = - \left( \psi_\mathrm{el} + \Delta\psi_\mathrm{ph} + \frac{J_\mathrm{e}}{3} \, {\rm tr} \bm\sigma \right) \bm{I},
\end{equation}
with ${\rm dev} ( \bm{T}) = \bm{T}- \frac{1}{3} \, {\rm tr}(\bm{T}) \, \bm{I}$. For an isotropic material, $\psi_\mathrm{el}$ can be expressed as a function of the principal invariants of $\bm \sigma$. Hence, the yield condition given by \cref{eq18} can now be formally formulated in $\bm \sigma$ only.

\subsection{Coupling to remodeling}
\label{S:2.3}
By remodeling, we understand the physiological replacement of one tissue by a different one. This effect can be described via our thermodynamic concept as well. For this purpose, two materials are  labeled by the index $i \in {1,2}$ are considered, which having total deformation gradients $\bm{F}_i$, inelastic deformation gradients $\bm{F}_{\mathrm{g}i}$, elastic energies $\psi_{\mathrm{el}i}$, physiological potentials $\Delta\psi_{\mathrm{ph}i}$ and dissipation potentials $\Delta_{\mathrm{g}i}$, as defined in the previous section. The formulation can be extended to an arbitrary number of tissues in a straightforward manner.

Assume, that the first tissue exists with the volume ratio $(1-\lambda)$ and the second one with the volume ratio $\lambda$. Moreover, suppose that both tissues have the same deformation gradient, i.e., $\bm{F}_1=\bm{F}_2=:\bm{F}$. This Taylor assumption is justified for tissues that are woven into each other and thus kinematically constrained - a situation typical for biological tissues. We obtain the total free energy
\begin{multline}
\label{eq25}
\psi(\bm{F},\bm{F}_{\mathrm{g}1},\bm{F}_{\mathrm{g}2},\lambda) = (1-\lambda) \, J_{\mathrm{g}1} \, \left( \psi_{\mathrm{el}1}(\bm{F} \cdot \bm{F}_{\mathrm{g}1}^{-1}) + \Delta\psi_{\mathrm{ph}1} \right) \\ + \lambda \, J_{\mathrm{g}2} \, \left( \psi_{\mathrm{el}2}(\bm{F} \cdot \bm{F}_{\mathrm{g}2}^{-1}) + \Delta\psi_{\mathrm{ph}2} \right).
\end{multline}

This allows us to define a thermodynamic driving force associated with remodeling as
\begin{equation}
\label{eq26}
q_\mathrm{rm} = - \frac{\partial \psi}{\partial \lambda} = J_{\mathrm{g}1} \, \left( \psi_{\mathrm{el}1}(\bm{F} \cdot \bm{F}_{\mathrm{g}1}^{-1}) + \Delta\psi_{\mathrm{ph}1} \right) - J_{\mathrm{g}2} \, \left( \psi_{\mathrm{el}2}(\bm{F} \cdot \bm{F}_{\mathrm{g}2}^{-1}) + \Delta\psi_{\mathrm{ph}2} \right).
\end{equation}

We see that without the presence of the physiological potentials stronger tissue with higher free energy would always be replaced by weaker tissue with lower free energy. Specifically, undamaged tissue would be replaced by damaged tissue. This would be contra-intuitive and contradict experimental observation in many cases.

To describe the evolution of the volume ration $\lambda$, we once again introduce a dissipation potential
\begin{equation}
\label{eq27}
\Delta_\mathrm{rm}(\dot{\lambda}) = r_\mathrm{rm} \, | \dot{\lambda} | + \frac{1}{2 {M^0_\mathrm{rm}}} \, \dot{\lambda}^2,
\end{equation}
allowing us to define a total dissipation potential {of the tissue} as
\begin{equation}
\label{eq28}
\Delta(\bm{L}_{\mathrm{g}1},\bm{L}_{\mathrm{g}2},\dot{\lambda}) = (1-\lambda) \, \Delta_{\mathrm{g}1}(\bm{L}_{\mathrm{g}1}) + \lambda \, \Delta_{\mathrm{g}2}(\bm{L}_{\mathrm{g}2}) + \Delta_\mathrm{rm}(\dot{\lambda}).
\end{equation}

Application of the {thermodynamic extremal} principle given in \cref{eq3} will return the evolution equation \cref{eq16} unchanged for $\bm{L}_{\mathrm{g}1}$ and $\bm{L}_{\mathrm{g}2}$, respectively. For the evolution of $\lambda$, we obtain
\begin{equation}
\label{eq29}
\dot{\lambda} ={M^0_\mathrm{rm}} \left( | q_\mathrm{rm} | - r_\mathrm{rm} \right)_+ \, \mathrm{sign}\left( q_\mathrm{rm} \right) .
\end{equation}

In most cases, we assume no threshold for the initiation of remodeling by setting $r_\mathrm{rm}=0$. Then \cref{eq29} takes the simple form
\begin{equation}
\label{eq30}
\dot{\lambda} = {M^0_\mathrm{rm}} \, q_\mathrm{rm}.
\end{equation}

In addition, in many cases, the mechanical properties of the completely healed tissue remain inferior to those of uninjured tissue \cite{frank1999optimisation,frank1999molecular}. Based on this experimental evidence, an irreversible stiffness loss parameter {$\eta$} is introduced, and the remodeling mobility is redefined as
\begin{equation}
\label{eq30b}
{M^0_\mathrm{rm} = M_\mathrm{rm}  \left(\eta -\lambda\right)_+,}
\end{equation}
yielding the evolution equation
\begin{equation}
\label{eq30a}
{\dot{\lambda} = M_\mathrm{rm} \, q_\mathrm{rm}\, \left(\eta -\lambda\right)_+.}
\end{equation}

Here, $\eta=0$ means that no damaged part can be healed and $\eta=1.0$ means that the material can be totally healed. 

\subsection{Coupling to damage}
\label{S:2.4}
The inclusion of material damage can now be easily accomplished. Let us assume material 1 to be the original tissue undergoing damage. Subsequently, it will be replaced by material 2 via remodeling during the healing process. Hence, material 2 represents ``scar tissue'' and is supposed to experience no further damage.

We introduce a constitutive function $f(d)$ describing the reduction in material stiffness, where $d$ is a damage parameter and {$f$} is at least twice differentiable, {monotonically decreasing} and satisfies the following conditions
\begin{equation}
\label{equ:5}
{f(0) = 1, \quad \mathop {\lim f(d) = 0}\limits_{d \to \infty }.}
\end{equation}

The elastic energy of material 1 is then premultiplied by $f(d)$, giving the new total energy
\begin{multline}
\label{eq31}
\psi(\bm{F},\bm{F}_{\mathrm{g}1},\bm{F}_{\mathrm{g}2},\lambda,d) = (1-\lambda) \, J_{\mathrm{g}1} \, \left( f(d) \, \psi_{\mathrm{el}1}(\bm{F} \cdot \bm{F}_{\mathrm{g}1}^{-1}) + \Delta\psi_{\mathrm{ph}1} \right) \\ + \lambda \, J_{\mathrm{g}2} \, \left( \psi_{\mathrm{el}2}(\bm{F} \cdot \bm{F}_{\mathrm{g}2}^{-1}) + \Delta\psi_{\mathrm{ph}2} \right).
\end{multline}

Let us study the energy in \cref{eq31} for a moment. With no physiological potentials present, the damaged material is the weaker one. Hence, thermodynamics would require undamaged material to be replaced with damaged material. To prevent this, we have to set $\Delta\psi_{\mathrm{ph}1}>0$. Since only the difference in the physiological potentials is relevant, we set $\Delta\psi_{\mathrm{ph}2}=0$. 
However, with this setting, material 1 is replaced with material 2 even when no damage has occurred. For this reason, the physiological potential has to become a time-dependent variable $\Delta\psi_{\mathrm{ph}1}:=\Delta\psi_\mathrm{ph}(t)$ and be connected to the evolution of $d$. This is possible only in a consistent way with respect to thermodynamics by declaring this variable to be an external one that can be influenced directly by the living body, for example, by sending messenger chemicals to the damaged tissue. We mimic this by introducing a dependence
\begin{equation}
\label{eq33a}
\Delta\psi_\mathrm{ph} = g(d),
\end{equation}	
where $g(d)$ is a positive, monotonically increasing function with $g(0)=0$. 

We obtain the thermodynamic driving force associated with damage as 
\begin{equation}
\label{eq32a}
q_\mathrm{d} = - \frac{\partial \psi}{\partial d} = - (1-\lambda) \, J_{\mathrm{g}1} \, f^\prime(d) \, \psi_{\mathrm{el}1}(\bm{F} \cdot \bm{F}_{\mathrm{g}1}^{-1}). 
\end{equation}

The dissipation potential associated with damage is introduced as
\begin{equation}
\label{eq34}
\Delta_\mathrm{d}(\dot{d}) = r_\mathrm{d} \, \dot{d} + \frac{1}{2 M_\mathrm{d}} \, \dot{d}^2, \qquad \dot{d}\geq 0.
\end{equation}

Note the constraint $\dot{d}\geq 0$ in \cref{eq34} prohibiting any reversal of damage. We obtain a new total dissipation potential of the form
\begin{equation}
\label{eq36}
\Delta(\bm{L}_{\mathrm{g}1},\bm{L}_{\mathrm{g}2},\dot{\lambda},\dot{d})  = (1-\lambda) \, \left( \Delta_{\mathrm{g}1}(\bm{L}_{\mathrm{g}1}) + J_{\mathrm{g}1} \, \Delta_\mathrm{d}(\dot{d}) \right) + \lambda \, \Delta_{\mathrm{g}2}(\bm{L}_{\mathrm{g}2}) + \Delta_\mathrm{rm}(\dot{\lambda}).
\end{equation}

We obtain the evolution equation of damage as
\begin{equation}
\label{eq37}
\dot{d} = M_\mathrm{d} \left( - f^\prime(d) \, \psi_{\mathrm{el}1}(\bm{F} \cdot \bm{F}_{\mathrm{g}1}^{-1}) - r_\mathrm{d} \right)_+.
\end{equation}

Note that $- f^\prime(d) \, \psi_{\mathrm{el}1}(\bm{F} \cdot \bm{F}_{\mathrm{g}1}^{-1}) \geq 0$.

\subsection{Summary of evolution equations in healing}
\label{S:2.51}
We summarize the obtained model in the following. Given ${\bm F}(t)$ we have evolution equations for the variables ${\bm F}_{\mathrm{g}1}(t)$, ${\bm F}_{\mathrm{g}2}(t)$, $\lambda(t)$ and $d(t)$ in the fully explicit form

\begin{tcolorbox}[ams gather]
	\label{eq39}
	\dot{\bm{F}}_{\mathrm{g}1} \cdot \bm{F}_{\mathrm{g}1}^{-1} = M_{\mathrm{g}1} \left( \| \tilde{\bm{q}}_{\mathrm{g}1} \| - r_{\mathrm{g}1} \right)_+ \, \mathrm{sign}\left( \tilde{\bm{q}}_{\mathrm{g}1} \right), \\
	\label{eq40}
	\dot{\bm{F}}_{\mathrm{g}2} \cdot \bm{F}_{\mathrm{g}2}^{-1} = M_{\mathrm{g}2} \left( \| \tilde{\bm{q}}_{\mathrm{g}2} \| - r_{\mathrm{g}2} \right)_+ \, \mathrm{sign}\left( \tilde{\bm{q}}_{\mathrm{g}2} \right), \\
	\label{eq41}
	\dot{\lambda} = M_\mathrm{rm} \left( | \tilde{q}_\mathrm{rm} | - r_\mathrm{rm} \right)_+ \, \mathrm{sign}\left( \tilde{q}_\mathrm{rm} \right), \\
	\label{eq42}
	\dot{d} = M_\mathrm{d} \left( \tilde{q}_\mathrm{d} - r_\mathrm{d} \right)_+.
\end{tcolorbox}

To account for the influence of damage, we introduce modified driving forces as follows:
\begin{tcolorbox}[ams gather]
	\label{eq44}
	\tilde{\bm{q}}_{\mathrm{g}1} = - \left( f(d) \, \psi_{\mathrm{el}1} + g(d) \right) \bm{I} + f(d) \, \bm{F}_{\mathrm{e}1}^\mathrm{T} \cdot \frac{\partial \psi_{\mathrm{el}1}}{\partial \bm{F}_{\mathrm{e}1}}, \\
	\label{eq45}
	\tilde{\bm{q}}_{\mathrm{g}2} = - \psi_{\mathrm{el}2} \, \bm{I} + \bm{F}_{\mathrm{e}2}^\mathrm{T} \cdot \frac{\partial \psi_{\mathrm{el}2}}{\partial \bm{F}_{\mathrm{e}2}},	\\
	\label{eq46}
	\tilde{q}_\mathrm{rm} = J_{\mathrm{g}1} \, \left( f(d) \, \psi_{\mathrm{el}1}(\bm{F} \cdot \bm{F}_{\mathrm{g}1}^{-1}) + g(d) \right) - J_{\mathrm{g}2} \, \psi_{\mathrm{el}2}(\bm{F} \cdot \bm{F}_{\mathrm{g}2}^{-1}), \\
	\label{eq47}
	\tilde{q}_\mathrm{d} = - f^\prime(d) \, \psi_{\mathrm{el}1}(\bm{F} \cdot \bm{F}_{\mathrm{g}1}^{-1}).
\end{tcolorbox}

\section{Nonlocal enhancement by gradient terms}
\label{S:2.6}
A nonlocal damage model is usually required because of the need to remove the pathological mesh dependence. Therefore, following the work of Dimitrijevic and Hackl \cite{dimitrijevic2008method,dimitrijevic2011regularization}, a gradient-enhanced nonlocal free energy term is added to the strain energy given in \cref{eq31}:

\begin{equation}
\label{eq32}
\begin{split}
\psi(\bm{F},\bm{F}_{\mathrm{g}1},\bm{F}_{\mathrm{g}2},\lambda,d) =& (1-\lambda)  J_{\mathrm{g}1}  ( f(d) \psi_{\mathrm{el}1}(\bm{F} \cdot \bm{F}_{\mathrm{g}1}^{-1})  +\frac{c_d}{2}{\| {{\nabla _{\bm{X}}}\phi } \|^2} + \frac{{{\beta _d}}}{2}{[ \phi  - {\gamma _d}d ]^2}+\\ &\Delta\psi_{\mathrm{ph}1})  + \lambda  J_{\mathrm{g}2}  ( \psi_{\mathrm{el}2}(\bm{F} \cdot \bm{F}_{\mathrm{g}2}^{-1}) + \Delta\psi_{\mathrm{ph}2} ).
\end{split}
\end{equation}

In \cref{eq32}, $c_d$ represents the gradient parameter that defines the degree of gradient regularization and the internal length scale. Three other variables are introduced as well:

- the field variable $\phi$, which {introduces an internal length scale via its gradient occurring in the expression for the energy}, 

- the energy-related penalty parameter $\beta_d$, which approximately enforces the local damage field and the nonlocal field to coincide, 

- parameter $\gamma_d$, which is used as a switch between the local and enhanced models.

In order to better monitor the damage and G\&R process, we introduce a healing parameter $H(d,t)$ as
\begin{equation}
\label{equ:5a}
H(d,t)=(1-\lambda)f(d)+\lambda.
\end{equation}

The healing parameter $H(d,t)$ takes values between 0 and 1. When $H(d,t)=0$, the tissue is completely damaged and its local stiffness is null, whereas when $H(d,t)=1$, the tissue is completely healed with newly produced tissue replacing the previously damaged one.

\section{Numerical implementation}

\subsection{Constitutive model}
\label{S:2.8}

For the undamaged part $\psi_0$, a neo-Hookean hyperelastic constitutive model \cite{nolan2014robust} is used. It is written as
\begin{equation}
\label{equ.15}
\begin{split}
{\psi_0}=\rm \frac 1 2 \mu_{0} \emph{J}_{\mathrm{e}}^{-2/3}(\emph I_{1{\mathrm{e}}}-3)+\rm \frac 1 2 \kappa_{0}(\emph{J}_{\mathrm{e}}-1)^2,
\end{split}
\end{equation}
where $\mu_{0}$ and $\kappa_{0}$ are the shear and bulk moduli of the soft isotropic matrix, respectively. $\emph I_{1{\mathrm{e}}}=tr(\bm C_\mathrm{e})$ is the first invariant of right Cauchy-Green tensor $\bm C_\mathrm{e}$. 

\subsection{Total potential energy and variational form}
\label{S:2.5}
The general total potential energy of the nonlocal damage model is
\begin{equation}
\label{equ.6}
\Pi  = \int\limits_\Omega  \psi \, \mathrm{d} V - \int\limits_\Omega  {\bar {\bm B}} \cdot {\bm{\varphi}} \, \mathrm{d} V - \int\limits_{\partial \Omega } {\bar {\bm T}} \cdot {\bm{\varphi}} \, \mathrm{d} S,
\end{equation}
where $\bar {\bm B}$ is the body force vector per unit reference volume of $\Omega$ and $\bar {\bm T}$ is the traction on the boundary $\partial \Omega$.

Minimization of the potential energy with respect to the primal variables $\bm \varphi$ and $\phi$ results in a coupled nonlinear system of equations that may be written as
\begin{equation}
\label{equ.7}
\int\limits_\Omega  {{\bm{P}}:{\nabla _{\bm{X}}}\delta {\bm{\varphi }}} \, \mathrm{d} V - \int\limits_\Omega  {\bar {\bm B}} \cdot \delta {\bm{\varphi }} \, \mathrm{d} V - \int\limits_{\partial \Omega } {\bar {\bm T} \cdot \delta {\bm{\varphi}}} \, \mathrm{d} S = 0,
\end{equation}

\begin{equation}
\label{equ.8}
\int\limits_\Omega  {{\bm{Y}}:{\nabla _{\bm{X}}}\delta \phi } \, \mathrm{d} V - \int\limits_\Omega  {Y\delta \phi } \, \mathrm{d} V = 0,
\end{equation}
where $\bm P$ is the first Piola-Kirchhoff stress, $\bm Y$ is vectorial damage quantity related to flux terms and $Y$ is the scalar damage quantity associated to source terms. They are defined as
\begin{equation}
\label{equ.9}
{\bm{P}} = {\partial _{{\bm F}}}\varPsi ,\quad \quad  \bm{Y}  = {\partial _{{\nabla _{\bm X}}\phi }}\varPsi ,\quad \quad
Y =- {\partial _\phi }\varPsi .
\end{equation}

The corresponding spatial quantities in \cref{equ.9} are given by
\begin{equation}
\label{equ.11}
{\bm{\sigma }} = {\bm{P}} \cdot {\rm{cof(}}{{\bm{F}}^{ - 1}}{\rm{)}},\quad \quad \ \ 
{\bar {\bm b}}=J^{-1}{\bar {\bm B}},
\end{equation}

\begin{equation}
\label{equ.12}
\bm y = {\bm{Y}} \cdot {\rm{cof(}}{{\bm{F}}^{ - 1}}{\rm{)}},\quad \quad \ \ 
y = {J^{ - 1}}Y,
\end{equation}
where ${\rm {cof}} ({\bm F})=J{\bm F}^{\rm -T}$.

Substituting \cref{equ.2,equ.9} into \cref{equ.7,equ.8}, the variational forms in the spatial description are
\begin{equation}
\label{equ.13}
\int\limits_\Omega  {{\bm{\sigma}}:{\nabla _{\bm{x}}}\delta {\bm{\varphi }}} \, \mathrm{d} v - \int\limits_\Omega  {\bar {\bm b}} \cdot \delta {\bm{\varphi }} \, \mathrm{d} v - \int\limits_{\partial \Omega } {\bar {\bm t} \cdot \delta {\bm{\varphi}}} \, \mathrm{d} s,
\end{equation}

\begin{equation}
\label{equ.14}
\int\limits_\Omega  {{\bm{y}}:{\nabla _{\bm{x}}}\delta \phi } \, \mathrm{d} v - \int\limits_\Omega  {y\delta \phi } \, \mathrm{d} v = 0.
\end{equation}

\subsubsection{Finite element discretization}
\label{S:2.3.1}
Isoparametric interpolations of the geometry variables $\bm X$, field variables $\bm \varphi$ and nonlocal field $\phi$ are respectively written as 
\begin{equation}
\label{equ:33}
{{\bm{X}}^h} = \sum\limits_{I = 1}^{n_{en}^{\bm{\varphi}} } {{N_I}\left( \xi  \right)} {{\bm{X}}_I},\quad 
{{\bm{\varphi }}^h} = \sum\limits_{I = 1}^{n_{en}^{\bm{\varphi}}} {{N_I}\left( \xi  \right)} {{\bm{\varphi }}_I},\quad 
{\phi ^h} = \sum\limits_{I = 1}^{n_{en}^\phi } {{N_I}\left( \xi  \right)} {\phi _I},
\end{equation}
where $\xi$ denotes the coordinates in the reference element, ${n_{en}^{\bm{\varphi}}}$ and ${n_{en}^{\phi}}$ are the nodal displacements and nodal nonlocal damage variables, respectively. 

The FE interpolations of \cref{equ:33} are introduced int the coupled nonlinear system of \cref{equ.13,equ.14}. To solve the coupled non-linear system of equations, an increment-iterative Newton-Raphson-type scheme is adopted: 
\begin{equation}
\label{equ:34}
{\left[ {\begin{array}{*{20}{c}}
		{{{\bm{R}}_{\bm{\varphi }}}}\\
		{{{\bm{R}}_\phi }}
		\end{array}} \right]^{i}}{ + }{\left[ {\begin{array}{*{20}{c}}
		{{{\bm{K}}_{{\bm{\varphi \varphi }}}}}\quad & {{{\bm{K}}_{{\bm{\varphi }}\phi }}}\\
		{{{\bm{K}}_{\phi {\bm{\varphi }}}}}\quad & {{{\bm{K}}_{\phi \phi }}}
		\end{array}} \right]^{i}} \cdot {\left[ {\begin{array}{*{20}{c}}
		{\Delta {\bm{\varphi }}}\\
		{\Delta \phi }
		\end{array}} \right]^{{i + 1}}}{ = }\left[ {\begin{array}{*{20}{c}}
	{\bf{0}}\\
	{\bf{0}}
	\end{array}} \right],,
\end{equation}
where 
\begin{equation}
\label{equ:35}
{{\bm{K}}_{{\bm{\varphi \varphi }}}}=\int_\Omega  {\nabla _{\bm {x}}^T N \cdot [ {{{\bm{C}}_h}({d},{t})} ] \cdot } {\nabla _{\bm x}}N {\rm{d}}v + \int_\Omega  {\left[ {\nabla _{\bm x}^T N \cdot {\bm{\sigma }}\cdot {\nabla _{\bm x}}N} \right] {\bm{I}}{\rm{d}}v},
\end{equation}

\begin{equation}
\label{equ:36}
{{\bm{K}}_{{\bm{\varphi }}\phi }}=\int_\Omega  {\nabla _{\bm x}^T N \cdot \frac{{{\rm d}{\bm{\sigma }}}}{{{\rm d}\phi }} \cdot N{\rm{d}}v} ,
\end{equation}

\begin{equation}
\label{equ:37}
{{\bm{K}}_{\phi {\bm{\varphi}}}}=\int_\Omega  {N_{}^T \cdot 2\frac{{{\rm d}y}}{{{\rm d}{\bf{g}}}} \cdot \nabla _{\bm x}^T N{\rm{d}}v} ,
\end{equation}

\begin{equation}
\label{equ:38}
{{\bm{K}}_{\phi \phi }}=\int_\Omega  {N_{}^T \cdot \frac{{{\rm d}y}}{{{\rm d}\phi }} \cdot N {\rm{d}}v}  +\int_\Omega  {\nabla _{\bm x}^T N \cdot \frac{{{\rm d}{\bm{y}}}}{{{\rm d}\phi }} \cdot \nabla _{\bm x}^T N{\rm{d}}v} .
\end{equation}

In the above equations the tangent terms ${{\rm d} {\bm{{\sigma}}}/{{\rm d} {\phi}}}$, $2{{\rm d} y}/{{\rm d} {\bf g}}$, ${{\rm d} y}/{{\rm d} {\phi}}$ and ${{\rm d} {\bm{y}}/{{\rm d} {\phi}}}$ are similar to the ones derived by Waffenschmidt et al. \cite{waffenschmidt2014gradient} and Polindara et al. \cite{polindara2017computational}, and ${{\bm{C}}_h}({d},{t})$ is a new time-dependent tangent stress-strain matrix in the damage and healing process given by
\begin{equation}
\label{equ:39}
{{\bm{C}}_h}({d},{t})=4f(d)(1-\lambda)\frac{\partial ^2 \psi_{\mathrm{el}1}(\bm C_e) }{\partial\bm C_e \partial \bm C_e}+
4\lambda \frac{\partial ^2 \psi_{\mathrm{el}2}(\bm C_e) }{\partial\bm C_e \partial \bm C_e},
\end{equation}
where the detailed expressions of $\frac{\partial ^2 \psi_{\mathrm{el}1}(\bm C_e) }{\partial\bm C_e \partial \bm C_e}$ and $\frac{\partial ^2 \psi_{\mathrm{el}2}(\bm C_e) }{\partial\bm C_e \partial \bm C_e}$ can be found in Nolan et al. \cite{nolan2014robust}.

\section{Numerical examples}
\label{S:3}
The model proposed in this paper is implemented within the commercial FE software Abaqus/Standard by means of a user subroutine UEL. Three numerical examples are shown onwards to illustrate the damage and healing effects in soft tissues with this model. In each example, an exponential damage function $f(d)=e^{-d}$ is adopted, but any other damage function satisfying \cref{equ:5} could be used. In both examples, the subscript 1 indicates the damaged tissues and 2 denotes the newly deposited part. Moreover, only the assumptions that the material properties of the newly deposited part are the same as the original tissues are considered in the following simulations.

\subsection{Uniaxial tension}
\label{S:5.1}
\begin{figure}[htbp]
	\subfigure[]{
		\begin{minipage}[t]{0.5\linewidth}
			\includegraphics[scale=0.8]{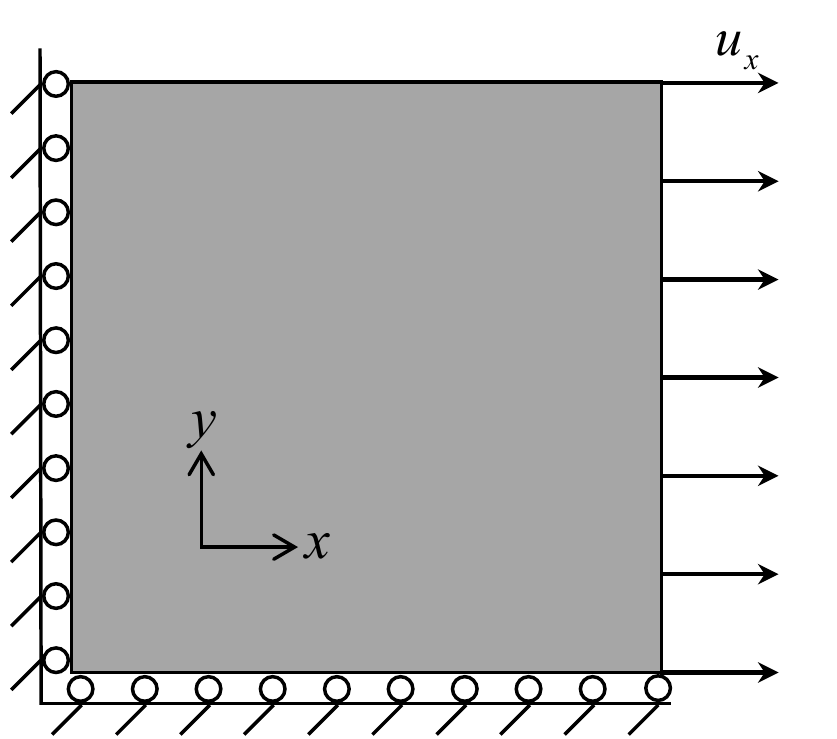}
			\label{fig:2a}
		\end{minipage}
	}
	\subfigure[]{
		\begin{minipage}[t]{0.5\linewidth}
			\includegraphics[scale=0.8]{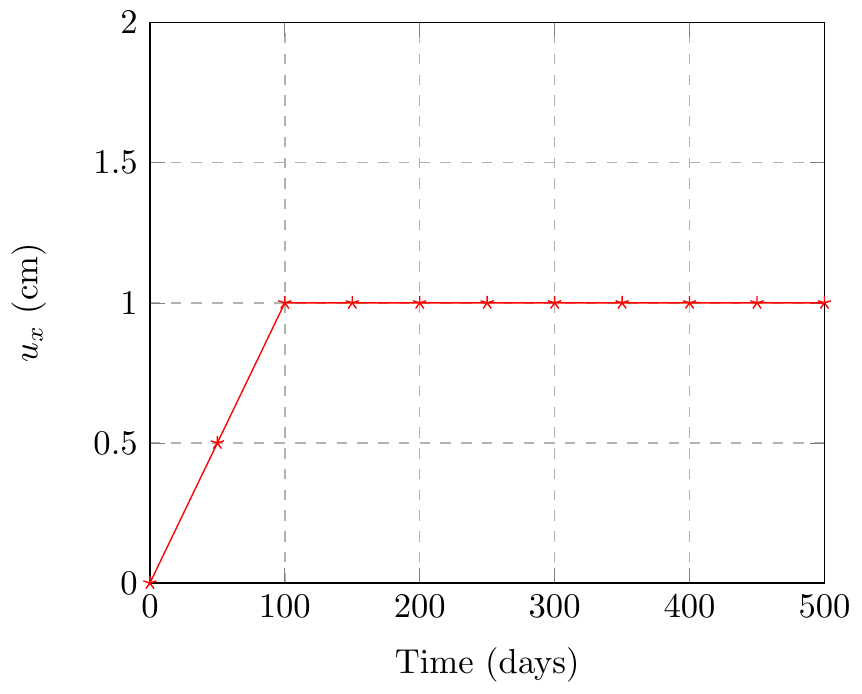}
			\label{fig:2b}
		\end{minipage}
	}
	\caption{Geometry and displacement used in the uniaxial tensile case study. (a) Geometric model, and (b) Loading curve.}
	\label{fig:ex.1}
\end{figure}

The first example is shown in \cref{fig:ex.1}. A square plate with a $10 \ \rm mm$ edge length is subjected to uniaxial tensile loading. As shown in \cref{fig:2b}, the displacement increases continuously from 0-100 days and is kept constant after the 100th day. The G\&R process is assumed to start from time t=100 days, and only one finite element is used in this example. The detailed material parameters are shown in \cref{tab:1-material}. 
The physiological potential of original tissues is set as $\Delta\psi_\mathrm{ph1}=0.001\ J$ in this example.

\begin{figure}[htbp]
	\subfigure[]{
		\begin{minipage}{0.5\linewidth}
			\centering
			\includegraphics[scale=0.8]{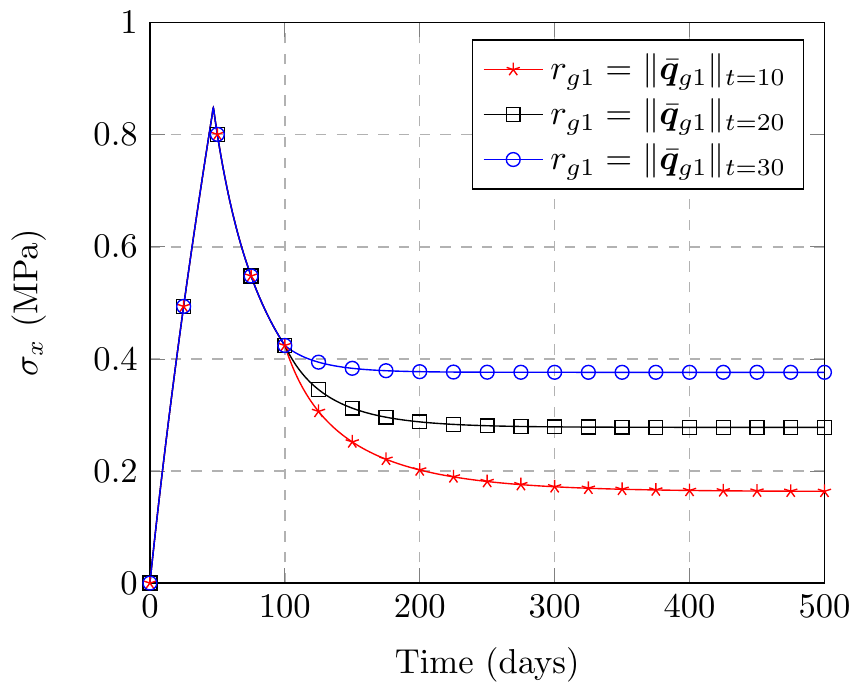}
			\label{fig:1-1-stress-different-rg}
		\end{minipage}
	}
	\subfigure[]{
		\begin{minipage}{0.5\linewidth}
			\centering
			\includegraphics[scale=0.8]{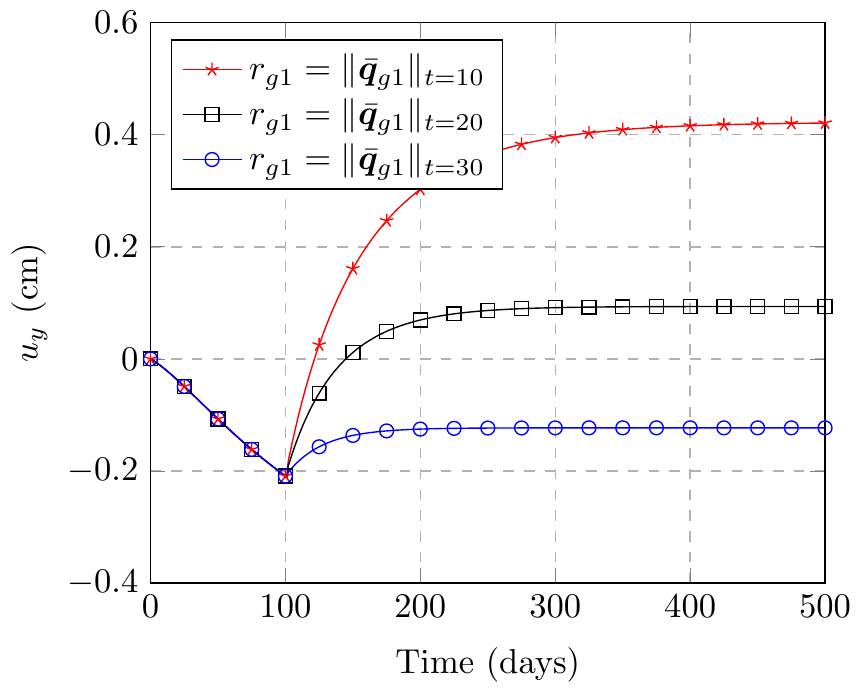}
			\label{fig:1-1-displacement-different-rg}
		\end{minipage}
	}
	\caption{Influence of the growth limit $r_{g1}$ in uniaxial tension. (a) The variation of the Cauchy stress $\sigma_x$ with time, and (b) the variation of the displacement $u_y$ with time.		
	}
\end{figure}

\begin{figure}[htbp]
	\subfigure[]{
		\begin{minipage}{0.5\linewidth}
			\centering
			\includegraphics[scale=0.8]{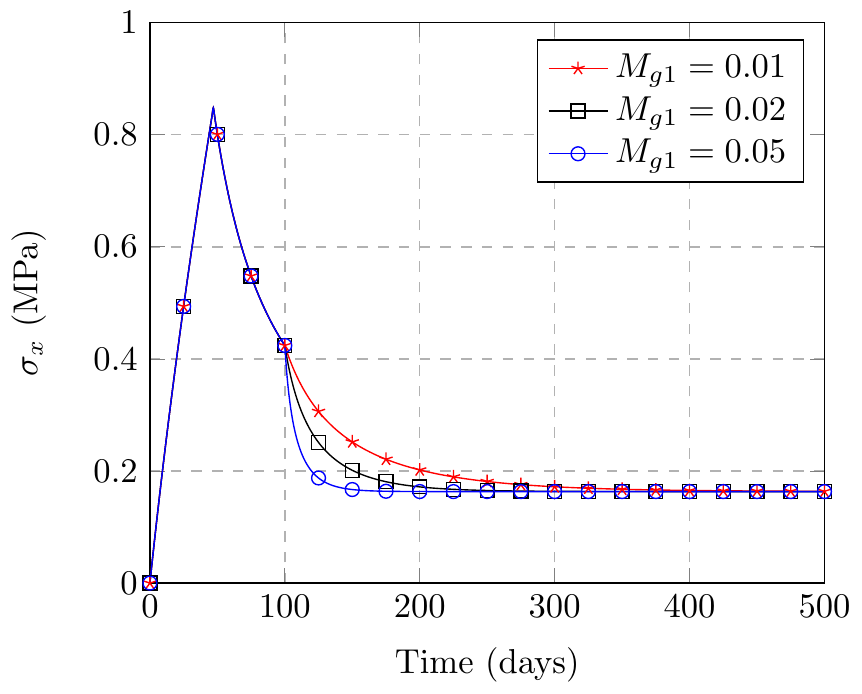}
			\label{fig:1-1-stress-different-mg}
		\end{minipage}
	}
	\subfigure[]{
		\begin{minipage}{0.5\linewidth}
			\centering
			\includegraphics[scale=0.8]{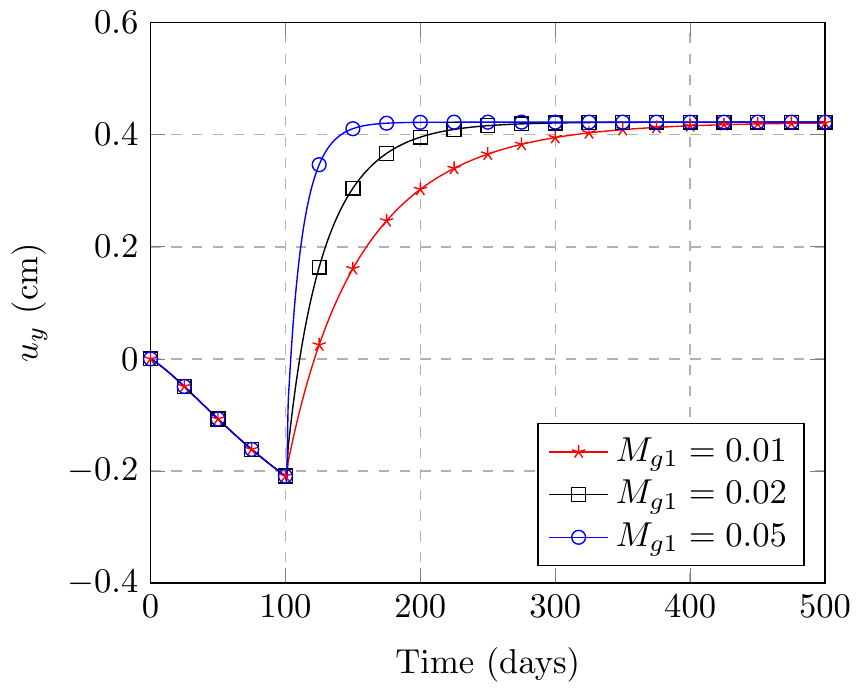}
			\label{fig:1-1-displacement-different-mg}
		\end{minipage}
	}
	\caption{Influence of the growth rate $M_{g1}$ in uniaxial tension. (a) The variation of the Cauchy stress $\sigma_x$ with time, and (b) the variation of the displacement $u_y$ with time.
	}
\end{figure}

Firstly, we check the the performance of the proposed model in simulating growth. Three values of growth limit $r_{g1}$ are set:
$r_{g1}=\|\bm q_{g1}\|_ {t=10\ { \rm days}}$, $r_{g1}=\|\bm q_{g1}\| _ {t=20\ { \rm days}}$ and $r_{g1}=\|\bm q_{g1}\| _ {t=50\ { \rm days}}$. The growth rate is $M_{g1}=0.01\ { \rm day^{-1}}$ and no remodeling is assumed to occur by setting $M_{rm}=0$. 
The variations of the Cauchy stress $\sigma_x$ and the displacement $u_y$ with time in \cref{fig:1-1-stress-different-rg,fig:1-1-displacement-different-rg} demonstrate that different homeostatic states can be reached by changing the growth limit $r_{g1}$, and a larger value of the $r_{g1}$ (in three values introduced above) leads to a larger homeostatic stress (see from \cref{fig:1-1-stress-different-rg}) and a smaller displacement (see $u_y$ from \cref{fig:1-1-displacement-different-rg}), it can be explained that a smaller gap between homeostatic and current state is produced by a larger $r_{g1}$, therefore, a smaller growth deformation is required in the healing process.
\cref{fig:1-1-stress-different-mg,fig:1-1-displacement-different-mg} show the influence of the growth rate $M_{g1}$ and it shows that a larger growth rate leads to a faster convergence of the homeostatic state.  

\begin{figure}[htbp]
	\subfigure[]{
		\begin{minipage}{0.5\linewidth}
			\centering
			\includegraphics[scale=0.8]{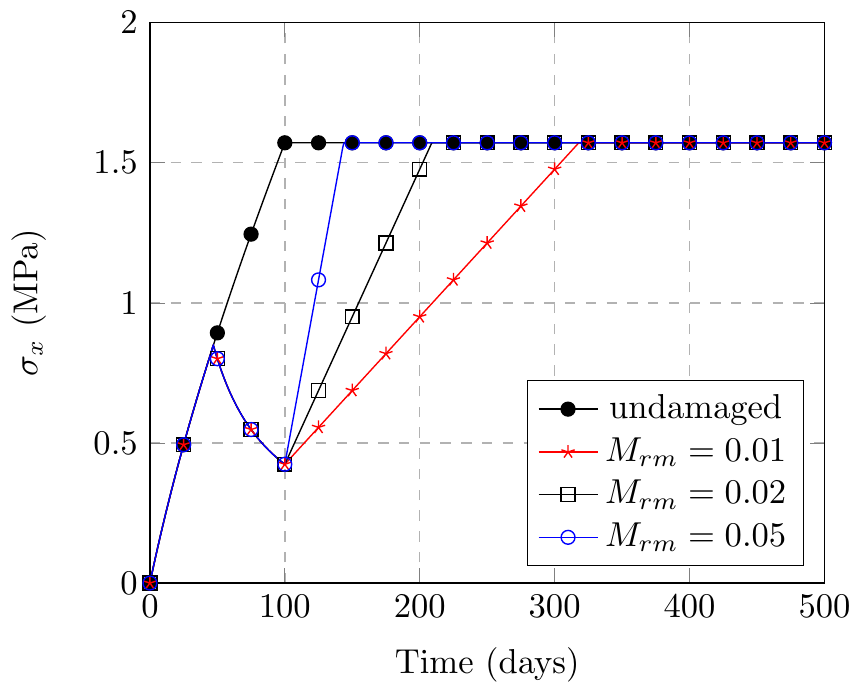}
			\label{fig:1-2-stress-different-rm}
		\end{minipage}
	}
	\subfigure[]{
		\begin{minipage}{0.5\linewidth}
			\centering
			\includegraphics[scale=0.8]{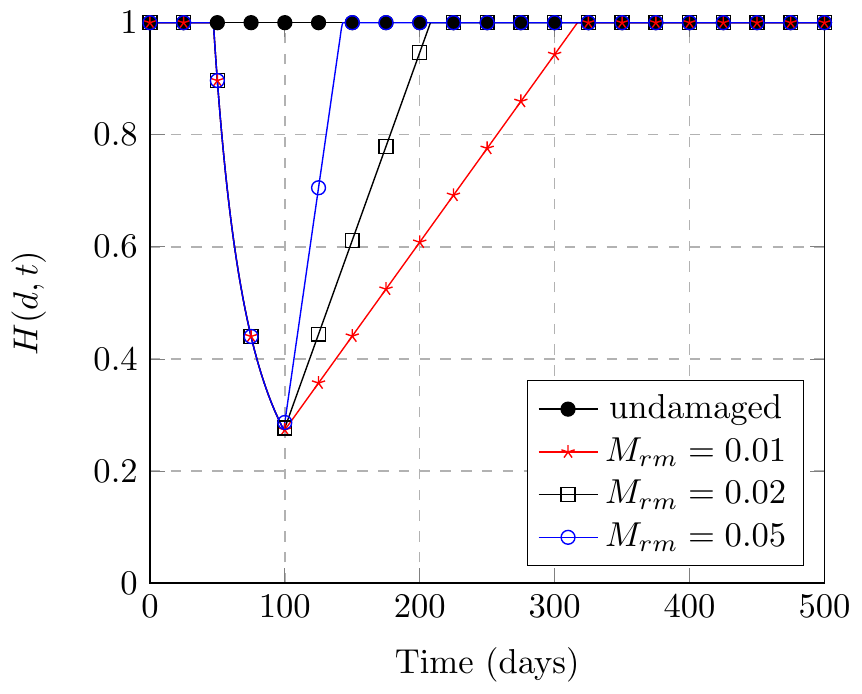}
			\label{fig:1-2-stress-different-lambda}
		\end{minipage}
	}
	\caption{Influence of the healing rate $M_{rm}$ in uniaxial tension. (a) The variation of the Cauchy stress $\sigma_x$ with time (b) the variation of the healing parameter $H(d,t)$ with time.}
	\label{fig:1-2-1}
\end{figure}

\begin{figure}[htbp]
	\subfigure[]{
		\begin{minipage}{0.5\linewidth}
			\centering
			\includegraphics[scale=0.8]{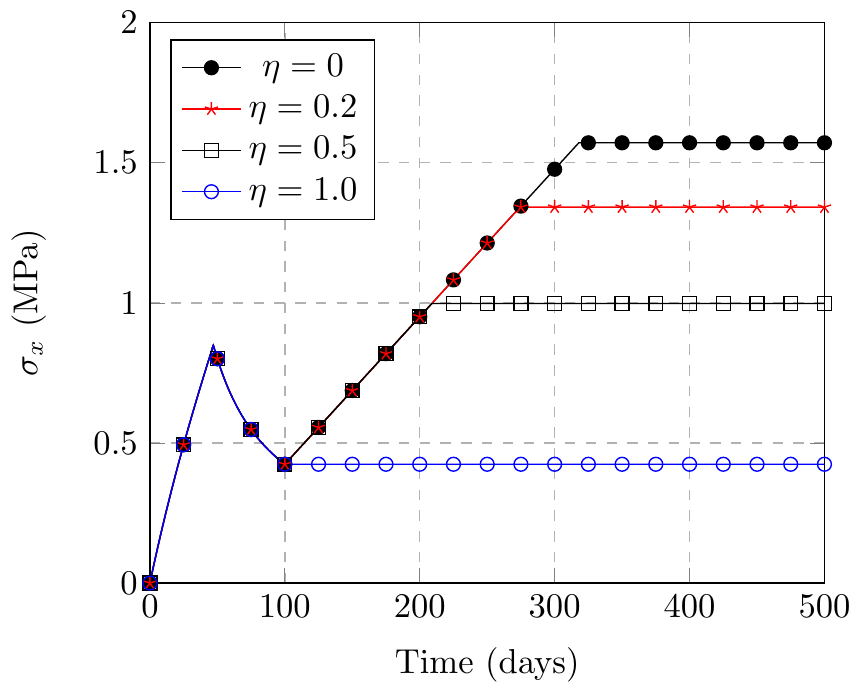}
			\label{fig:1-2-fd-different-rm}
		\end{minipage}
	}
	\subfigure[]{
		\begin{minipage}{0.5\linewidth}
			\centering
			\includegraphics[scale=0.8]{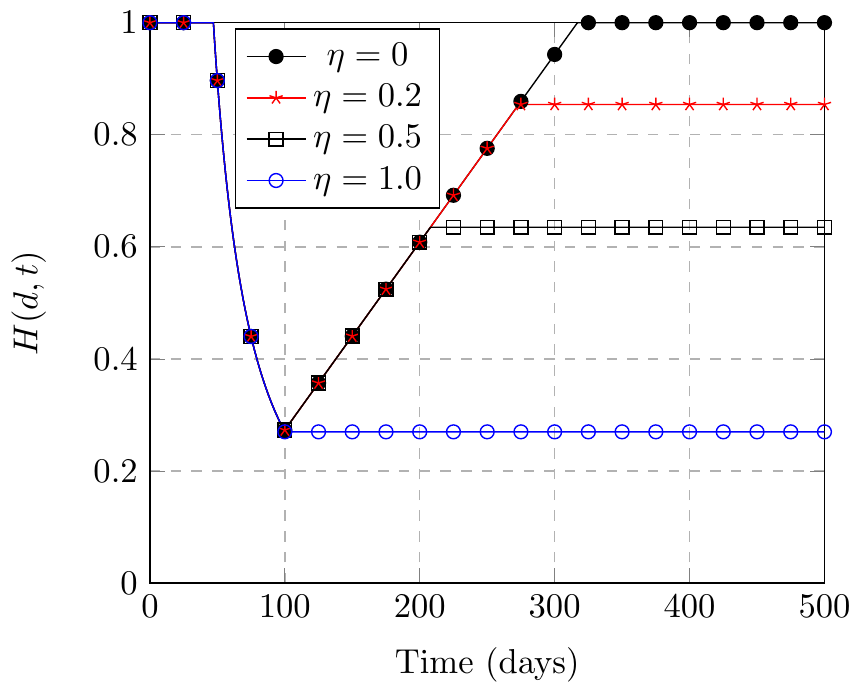}
			\label{fig:1-2-fd-different-lambda}
		\end{minipage}
	}
	\caption{Influence of the irreversible stiffness loss $\eta$ in uniaxial tension. (a) The variation of the Cauchy stress $\sigma_x$ with time, and (b) the variation of the healing parameter $H(d,t)$ with time.}
	\label{fig:1-2-2}
\end{figure}

Secondly, the performance of the proposed model in simulating remodeling without growth are shown in \cref{fig:1-2-1,fig:1-2-2}. 
Three different values of the rate of remodeling $M_{rm}$ are tested to check the influence of the remodeling rate $M_{rm}$ with the irreversible stiffness loss $\eta=0$. The results of the variations of $\sigma_x$ and $H(d,t)$ with time shown in \cref{fig:1-2-stress-different-rm,fig:1-2-stress-different-lambda} illustrate that a higher value of the remodeling rate $M_{rm}$ induces a faster remodeling speed, which means a shorter time is needed in the process of replacement of the damaged tissues by the newly deposited part.
The influence of the irreversible stiffness loss $\eta$ on the variations of $\sigma_x$ and $H(d,t)$ with time are shown in \cref{fig:1-2-fd-different-rm,fig:1-2-fd-different-lambda} by setting four different values, i.e., $\eta=0$, $\eta=0.2$, $\eta=0.5$ and $\eta=1.0$, when $M_{rm}=0.01 \ \rm days^{-1}$,
it is seen that the converged healing parameter $H(d,t)$ gradually increases with the decrease of $\eta$, and $H(d,t)$ converges to 1 when $\eta=0$ indicating a complete healing for soft tissue, while $H(d,t)$ converges to 0 meaning no healing occurs when $\eta=1$.

\begin{figure}[htbp]
	\subfigure[]{
		\begin{minipage}{0.5\linewidth}
			\centering
			\includegraphics[scale=0.8]{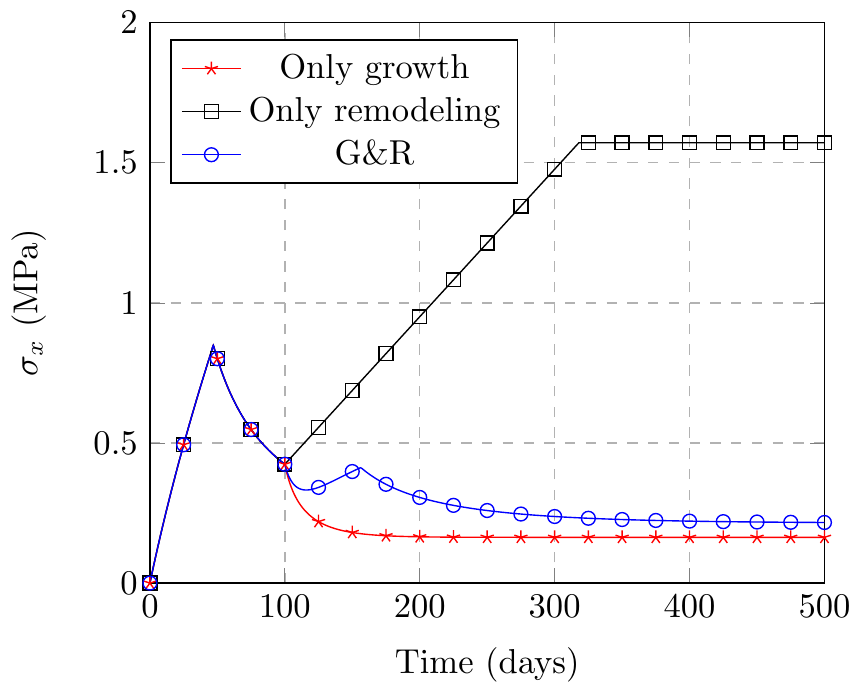}
			\label{fig:1-3-stress}
		\end{minipage}
	}
	\subfigure[]{
		\begin{minipage}{0.5\linewidth}
			\centering
			\includegraphics[scale=0.8]{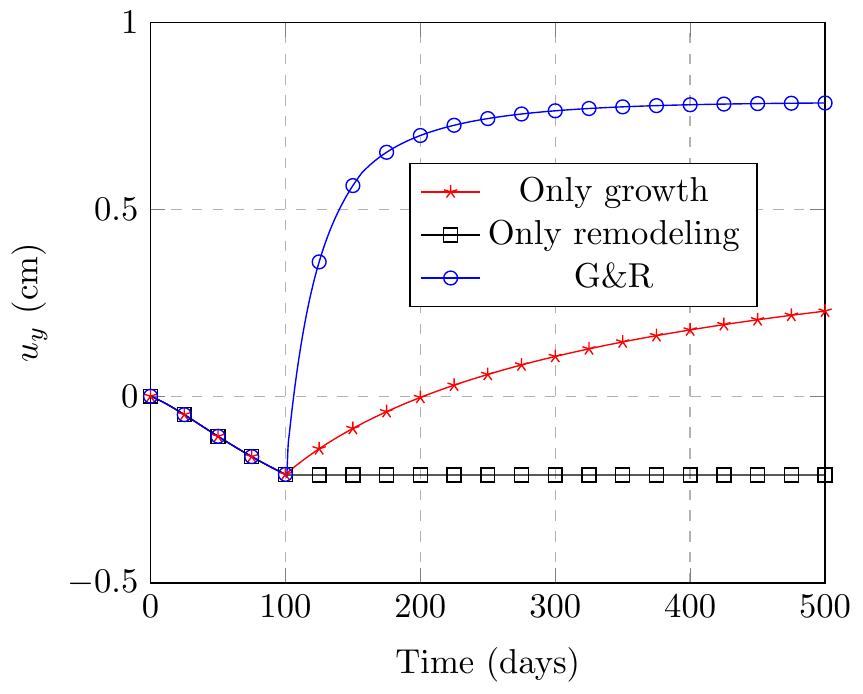}
			\label{fig:1-3-displacement}
		\end{minipage}
	}
	\caption{Results for the uniaxial tensile test with different situations. (a) The variation of the Cauchy stress $\sigma_x$ with time, and (b) the variation of the healing parameter $H(d,t)$ with time.}
	\label{fig:1-3}
\end{figure}

Thirdly, the combined effects of growth and remodeling are shown in \cref{fig:1-3}.
The G\&R parameters are set as follows: $r_{gi}=\|(\bm q_{gi})\|_ {t=10\ { \rm days}}$, $M_{g1}=M_{g2}=0.03\ \rm days^{-1}$, $M_{rm}=0.01 \ \rm days^{-1}$ and $\eta=0$. 
In \cref{fig:1-3-stress,fig:1-3-displacement}, the variations in $\sigma_x$ and $u_y$ with time are shown by comparing three cases: (1) Only growth occurs; (2) only remodeling occurs; and (3) growth and remodeling occur at the same time.
The result in \cref{fig:1-3-stress} illustrating that, when growth and remodeling combined occur, the stress decreases firstly due to the change in the configuration caused by growth and with the remodeling of the damaged tissues, and the stress increases until all damaged tissues are changed into the newly deposited part, and finally, the stress converges to a different homeostatic stress compared to the situation where only growth occurs.
Although remodeling does not change the homeostatic stress much (see from \cref{fig:1-3-stress}), a larger $u_y$ can be found in \cref{fig:1-3-displacement} by comparing with case (1) and (3), and that can be illustrated by the combined effects of G\&R such that a larger deformation is needed for the newly deposited part to converge to the homeostatic state than for the damaged part. 

\subsection{Open-hole plate}
\begin{figure}[htbp]
	\subfigure[]{
		\begin{minipage}[t]{0.5\linewidth}
			\includegraphics[scale=0.47]{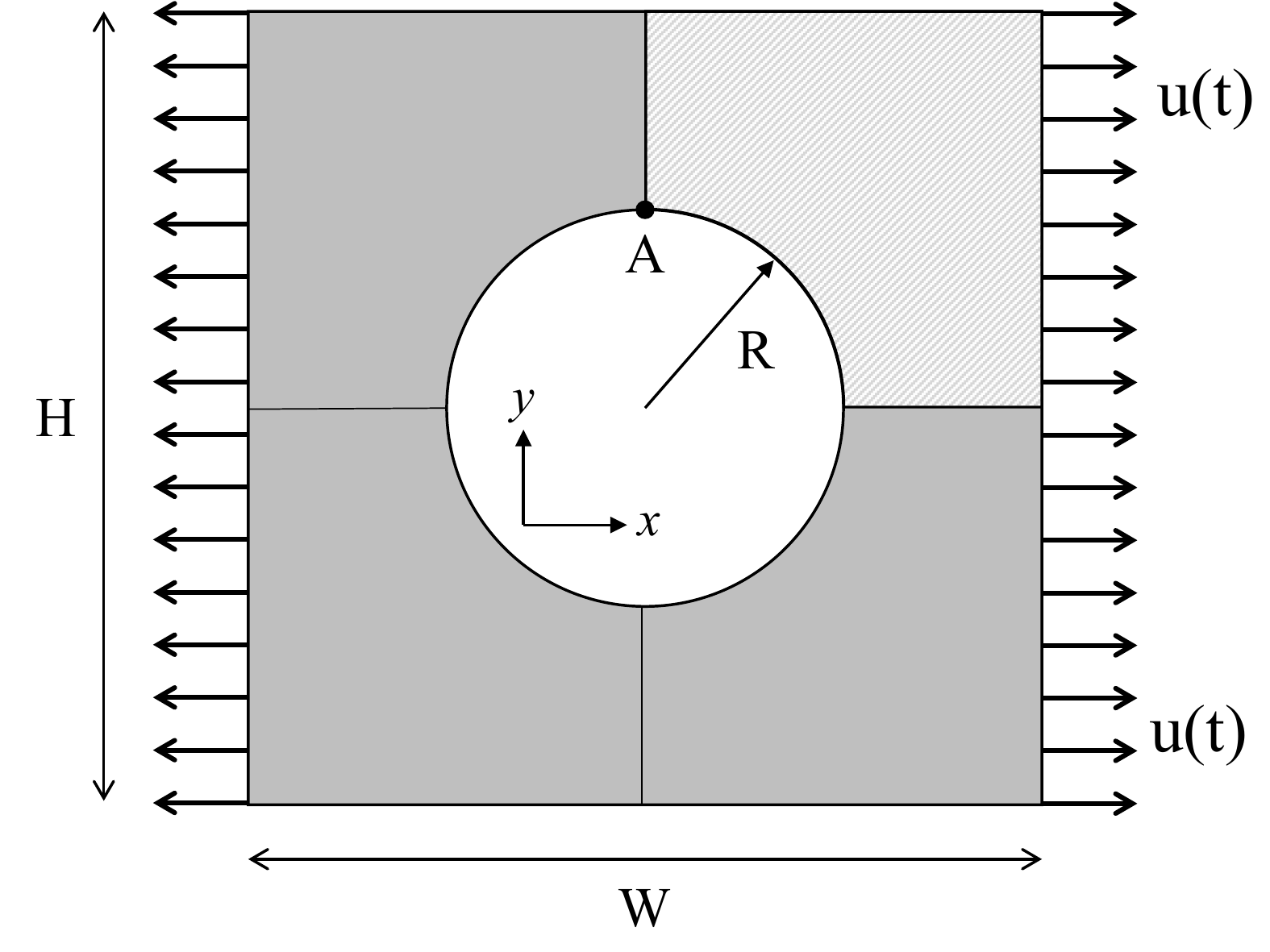}
			\label{fig:2-geo}
		\end{minipage}
	}
	\subfigure[]{
		\begin{minipage}[t]{0.5\linewidth}
			\includegraphics[scale=0.8]{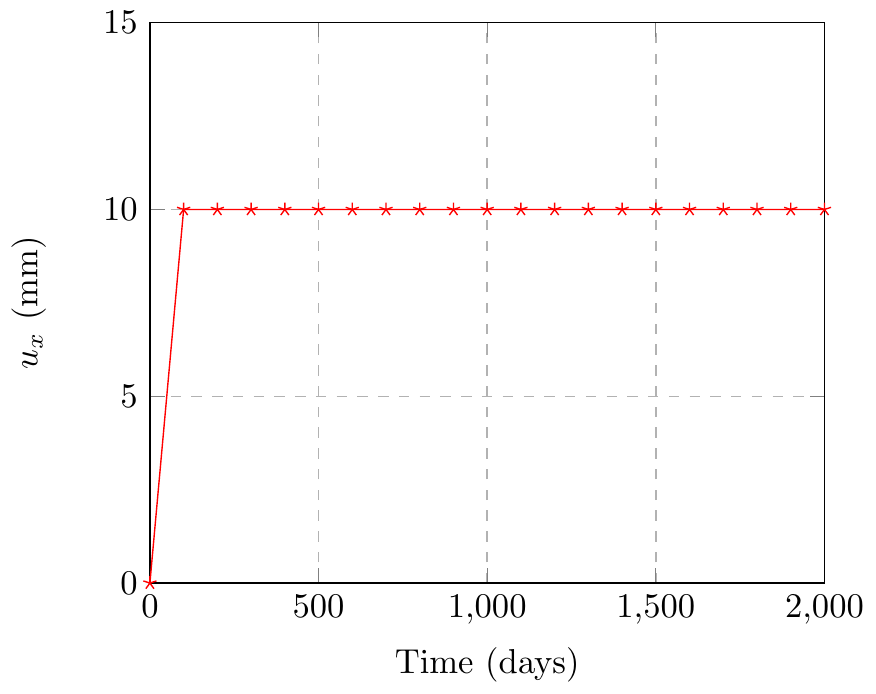}
			\label{fig:c-loading}
		\end{minipage}
	}
	\caption{Geometry and loading condition for the open-hole plate case. (a) Geometry, (b) loading curve.}
	\label{fig:ex.2}
\end{figure}

The second numerical example is an open-hole plate subjected to displacement loading. The geometry and the loading curve are shown in \cref{fig:ex.2}. The detailed material parameters are reported in \cref{tab:2-material}. Due to the symmetry, only 1/4 of the plate at the top right corner is analyzed. For all simulations, the irreversible stiffness loss parameter $\eta=0$.

\begin{figure}[htbp]
	\centering
	\includegraphics[scale=1.0]{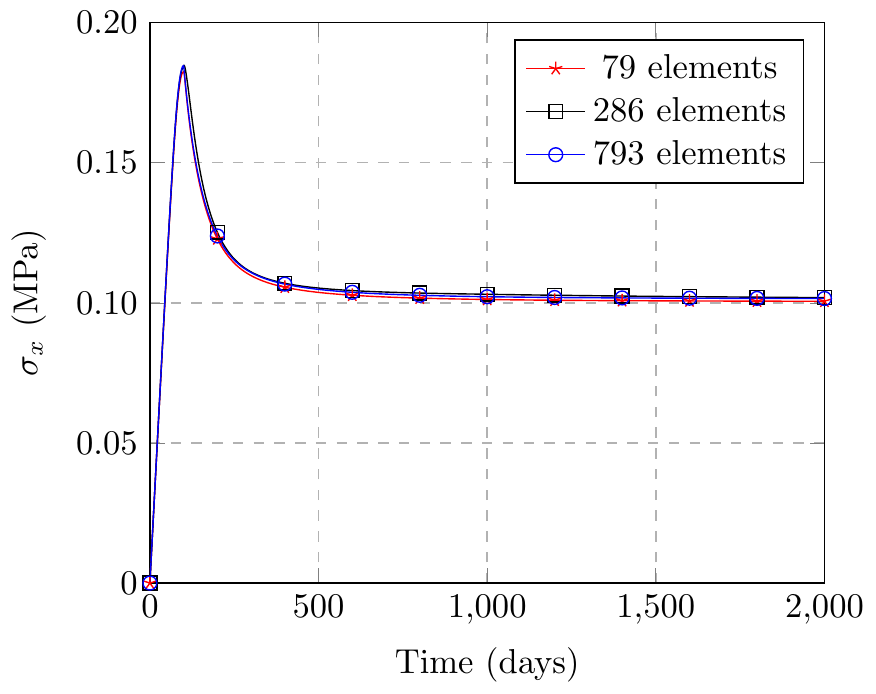}
	\caption{The average Cauchy stress $ \sigma_x$ of the right-hand side with different mesh sizes.}
	\label{fig:2-different-mesh}
\end{figure}

\begin{figure}[htbp]
	\centering
	\includegraphics[scale=0.35]{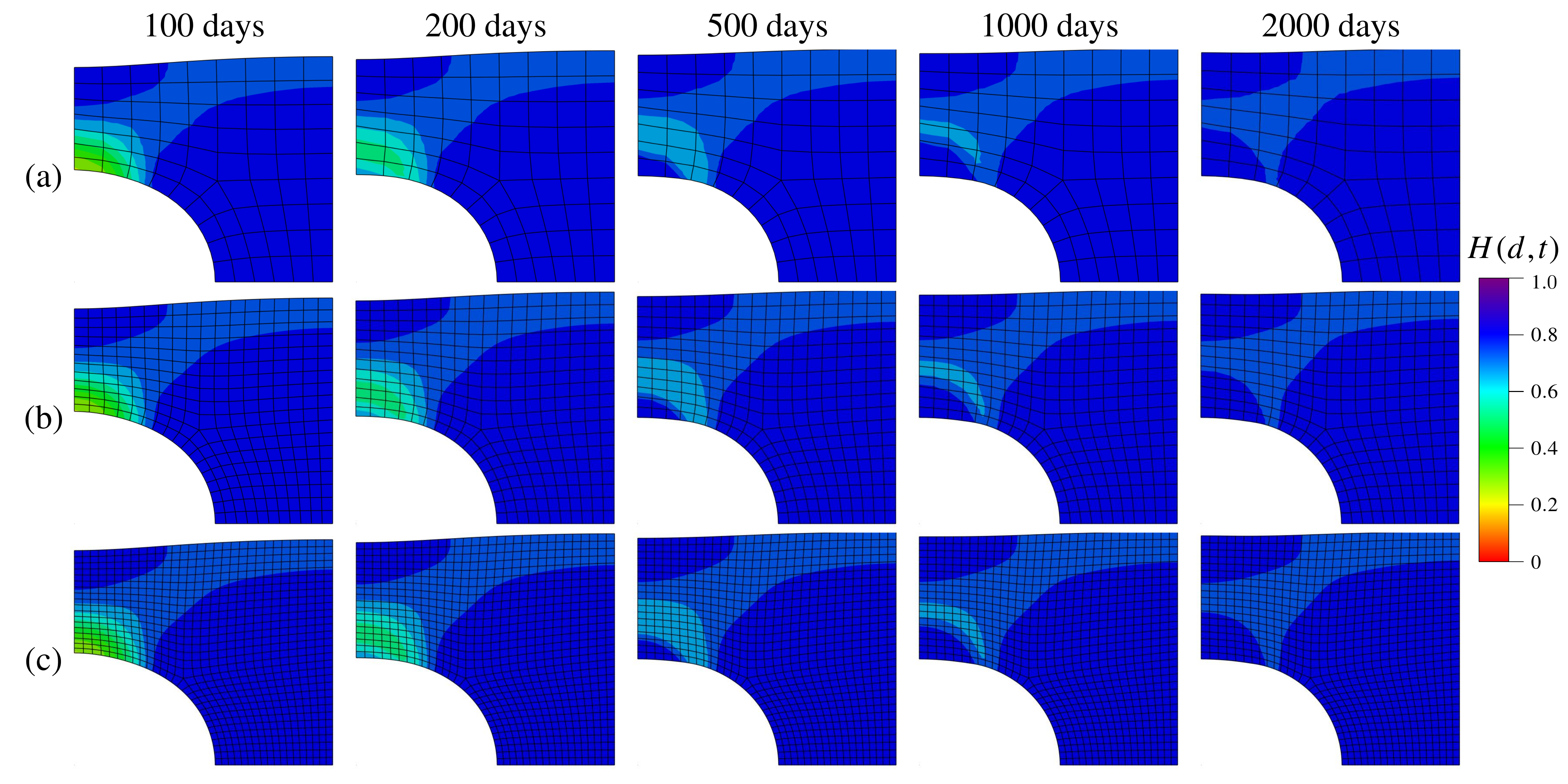}
	\caption{Variation of contours of the healing parameter $H(d,t)$ with time for different numbers of elements (a) 79 elements, (b) 286 elements, and (c) 793 elements.}
	\label{fig:2-fd-contours-different-mesh}
\end{figure}

Firstly, the mesh dependence of the proposed method is investigated by simulating three different mesh sizes (79 elements, 286 elements, and 793 elements), and the average Cauchy stress $\sigma_x$ of the right side and the contours for the healing parameter $H(d,t)$ at different time are shown in \cref{fig:2-different-mesh} and \cref{fig:2-fd-contours-different-mesh}, respectively.
The G\&R parameters are set as $M_{gi}=0.03 \ \rm days^{-1}$, $M_{rm}=0.1 \ \rm days^{-1}$ and $r_{gi}=\| (\bm q_{gi})\|_ {t=50\ { \rm days}}$, respectively. Both the stress curves and contours of the healing parameter illustrate that the results are rather similar for all different elements used, and a good mesh-independence is achieved by the proposed model.

\begin{figure}[htbp]
	\subfigure[]{
		\begin{minipage}{0.5\linewidth}
			\centering
			\includegraphics[scale=0.8]{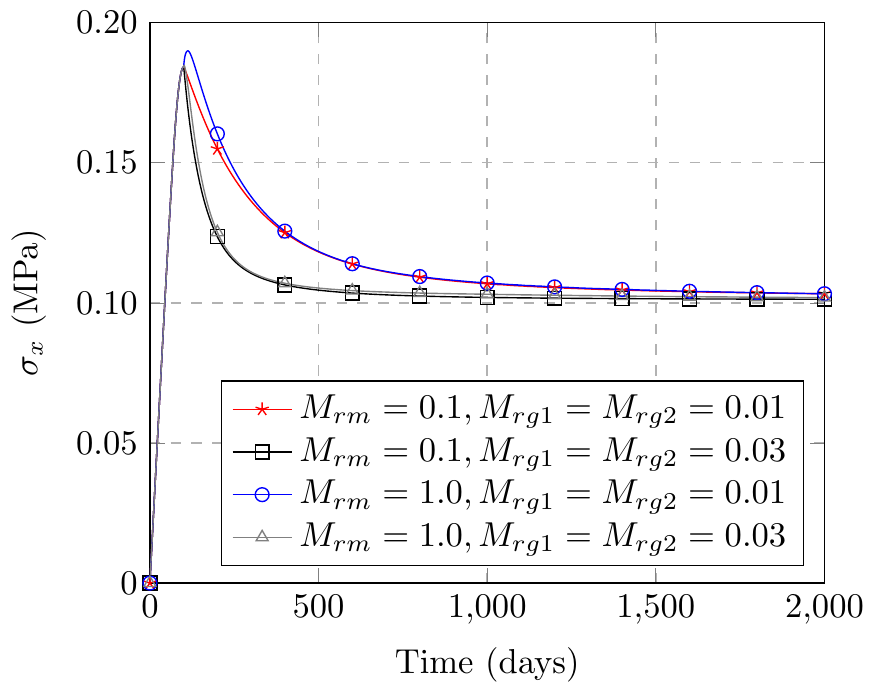}
			\label{fig:2-stress-different-parameter}
		\end{minipage}
	}
	\subfigure[]{
		\begin{minipage}{0.5\linewidth}
			\centering
			\includegraphics[scale=0.8]{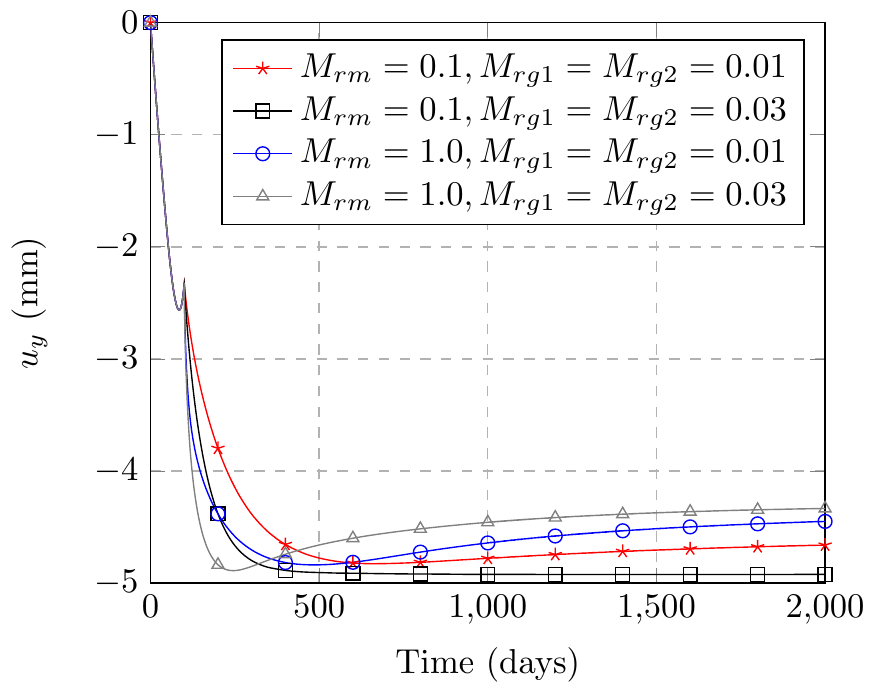}
			\label{fig:2-displacement-different-parameter}
		\end{minipage}
	}
	\caption{Influence of the rate of G\&R ($M_{rm}$, $M_{g1}$ and $M_{g2}$) in open-hole plate. (a) The variation of the Cauchy stress $\sigma_x$ with time, and (b) the variation of the displacement $u_y$ with time at node A.}
	\label{fig:2-different-parameter}
\end{figure}

Secondly, the influence of different G\&R parameters are analyzed. 
The sensitivity of parameters of G\&R $M_{rm}$, $M_{g1}$ and $M_{g2}$ are set as four different values with the growth limit $r_{gi}=\| (\bm q_{gi})\|_ {t=50\ { \rm days}}$.
It can be seen from \cref{fig:2-stress-different-parameter} that the Cauchy stress $\sigma_x$ can converge to the homeostatic state for all four cases, and a larger remodeling rate $M_{rm}$ induces a higher stress in the healing process when the growth rate $M_{gi}$ is same. 
Although the remodeling rate $M_{rm}$ has relatively less influence on the homeostatic stress (see from \cref{fig:2-stress-different-parameter}), the long-term evolution of the deformation is still depended on the remodeling rate $M_{rm}$, which can bee seen from \cref{fig:2-displacement-different-parameter} that the displacement at node A (the location is shown in \cref{fig:2-geo}) is shown. 
It is seen in \cref{fig:2-displacement-different-parameter} that, at the beginning of G\&R, a faster decrease of the displacement is caused by a higher growth rate,
and as for the homeostatic state, a higher remodeling rate $M_{rm}$ leads to a larger deformation $u_y$ with the same $M_{gi}$, while a larger deformation $u_y$ is produced with a smaller $M_{gi}$ when $M_{rm}$ is the same. 

\begin{figure}[htbp]
	\subfigure[]{
		\begin{minipage}{0.5\linewidth}
			\centering
			\includegraphics[scale=0.8]{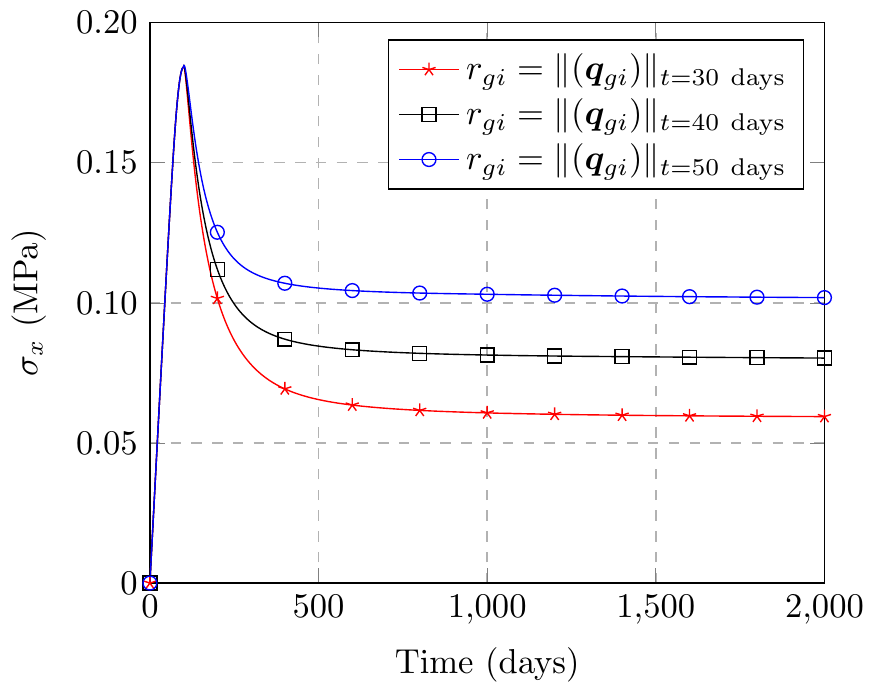}
			\label{fig:2-stress-different-rg}
		\end{minipage}
	}
	\subfigure[]{
		\begin{minipage}{0.5\linewidth}
			\centering
			\includegraphics[scale=0.8]{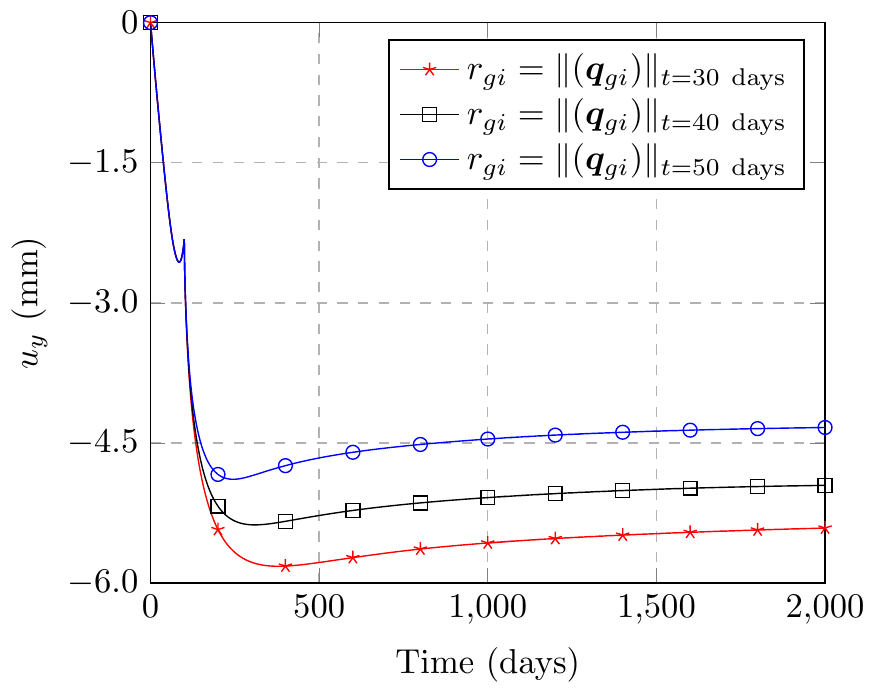}
			\label{fig:2-displacement-different-rg}
		\end{minipage}
	}
	\caption{Influence of the growth limit $r_{gi}$ in open-hole plate. (a) The variation of the Cauchy stress $\sigma_x$ with time, and (b) the variation of the displacement $u_y$ with time at node A.}
\end{figure}

The influence of growth limit is also investigated under three different values, i.e., $r_{gi}=\| (\bm q_{gi})\|_ {t=30\ { \rm days}}$, $r_{gi}=\| (\bm q_{gi})\|_ {t=40\ { \rm days}}$ and $r_{gi}=\| (\bm q_{gi})\|_ {t=50\ { \rm days}}$, and the average Cauchy stress $\sigma_x$ of the right side and the displacement $u_y$ at node A (the location is shown in \cref{fig:2-geo}) are shown in \cref{fig:2-stress-different-rg} and \cref{fig:2-displacement-different-rg}, respectively. The G\&R rate are set as $M_{gi}=0.03 \ \rm days^{-1}$ and $M_{rm}=1.0 \ \rm days^{-1}$, respectively. 
All three different $r_{gi}$ can converge to the homeostatic state, and the homeostatic stress increases with increasing the growth limit. 
Combining \cref{fig:2-stress-different-rg,fig:2-displacement-different-rg}, it can be found that a smaller displacement is produced when there exists a smaller gap between the current state and the homeostatic state to recover.

\begin{figure}[ht]
	\centering
	\includegraphics[scale=0.38]{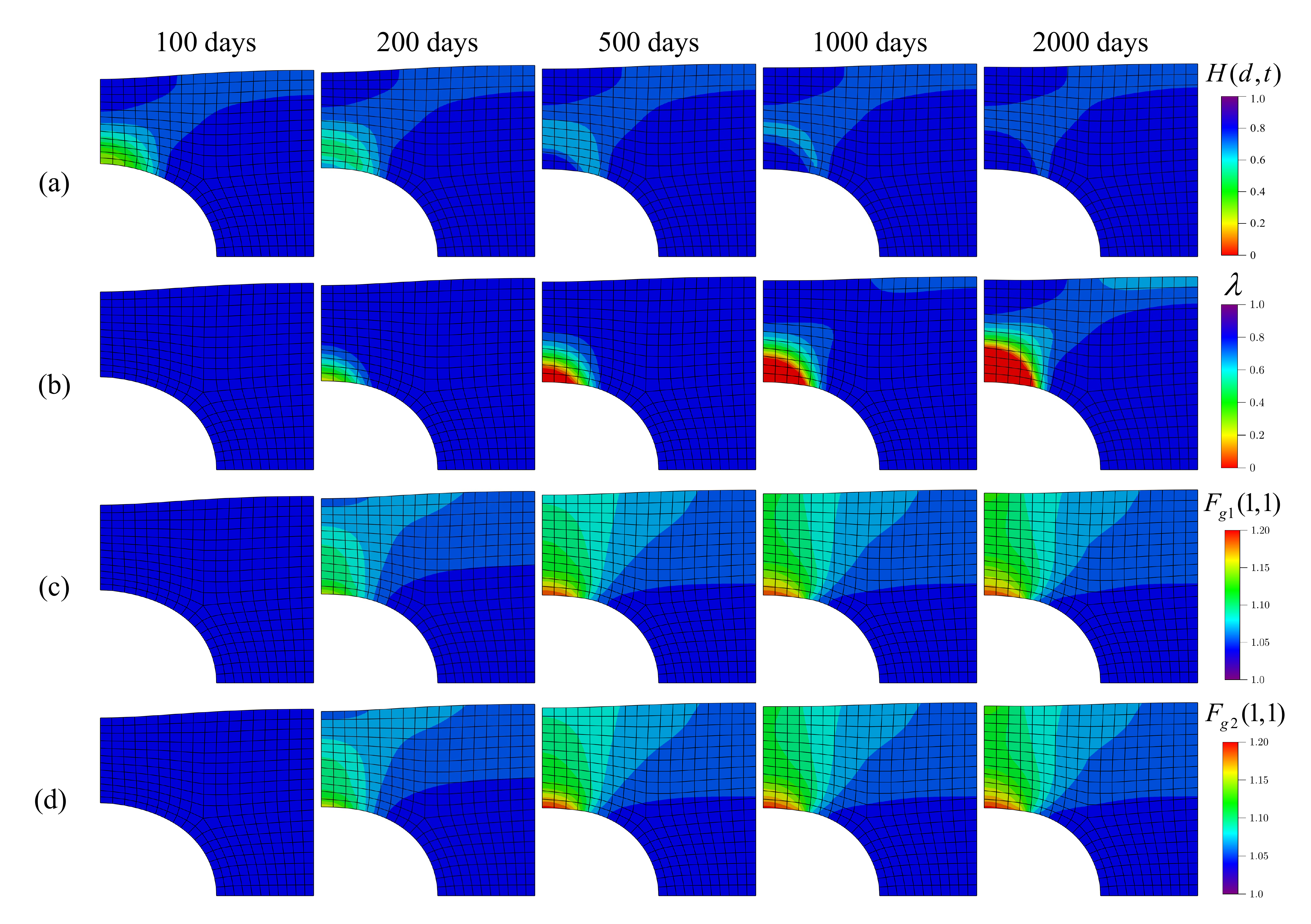}
	\caption{The contours for different parameters in healing process at different times. (a) The healing parameter $H(d,t)$, (b) the newly deposited part $\lambda$, (c) the component of the growth deformation for damaged part $F_{g1}(1,1)$, and (d) the component of the growth deformation for newly deposited part $F_{g2}(1,1)$.}
	\label{fig:2-different-parameter-contour}
\end{figure}

Thirdly, the evolution of the contours of the healing parameter $H(d,t)$, the volume ratio of the newly deposited part $\lambda$, and the component of the growth deformation for the damaged part $F_{g1}(1,1)$ and the newly deposited part $F_{g2}(1,1)$ through the healing process are shown in
Fig. 12(a)-(d), respectively, when $M_{gi}=0.03 \ \rm days^{-1}$, $M_{rm}=0.1 \ \rm days^{-1}$ and $r_{gi}=\| (\bm q_{gi})\|_ {t=50\ { \rm days}}$.
It can be observed that growth mainly occurs before the 500th day in this example, while remodeling occours for a relatively longer time, which is similar to the results shown in \cref{fig:2-different-parameter}.
The contours shown in \cref{fig:2-different-parameter-contour} exhibit the ability of our proposed model in predicting the evolution of G\&R over the long-term time again.

\subsection{Balloon angioplasty in atherosclerotic  artery}
\begin{figure}[htbp]
	\includegraphics[scale=0.7]{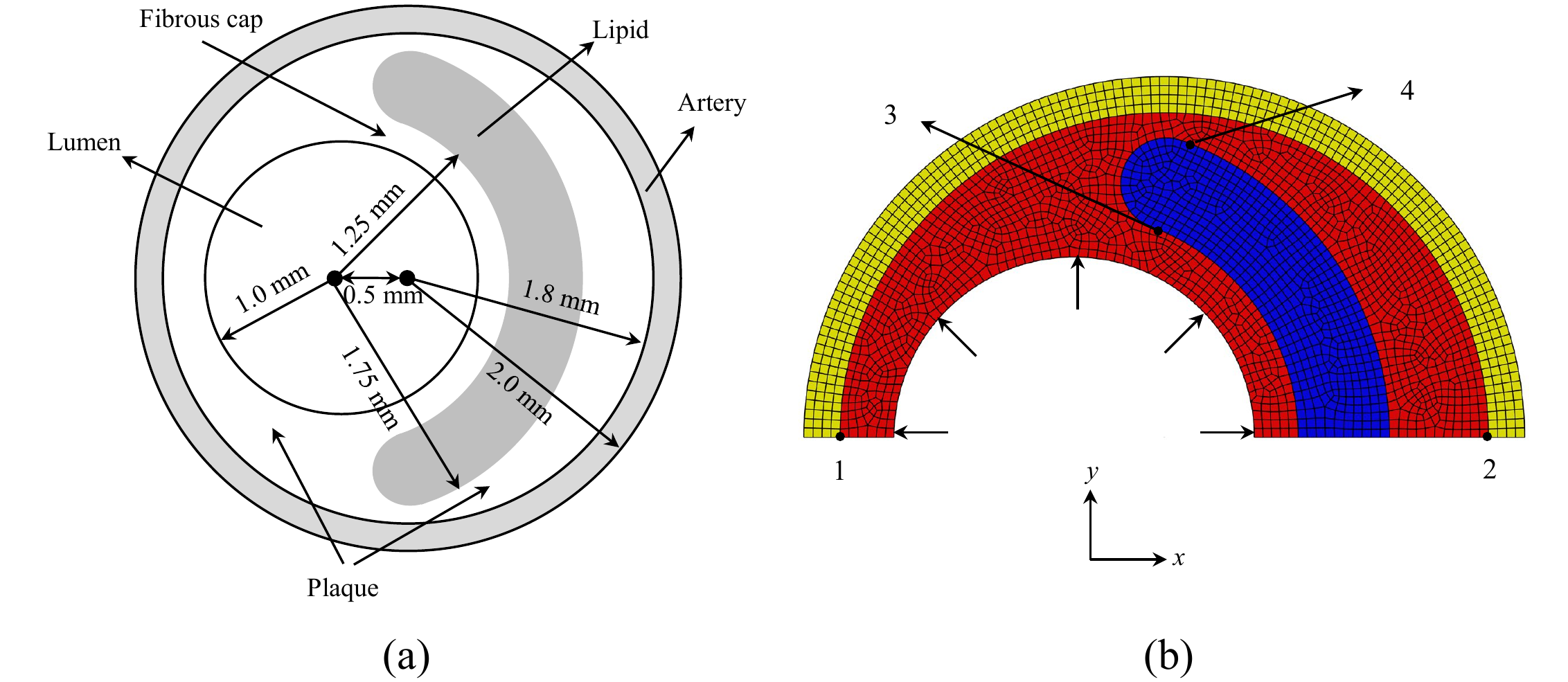}
	\caption{Geometric and FEM mesh model for balloon angioplasty.(a) Geometric model, and (b) FEM mesh model.}
	\label{fig:ex.3}
\end{figure}

The third example is associated with intraoperative injury and the long-term healing of atherosclerotic patients. The idealized two-dimensional cross-sectional model shown in \cref{fig:ex.3} was established by Loree et al. \cite{loree1992effects}.
The artery is modeled as a thick-walled cylinder with an inner radius of $1.8$ mm and an outer radius of $2.0$ mm. The lumen is modeled as a circular hole of radius $1.0$ mm with an eccentricity of $0.5$ mm with respect to the artery center. Fibrous plaque occupies the region between the luminal wall and the inner wall of the artery. The fibrous cap is assumed as continuous with the fibrous plaque and has the same material properties as the fibrous plaque.
A subintimal lipid pool exists as a $140^o$ crescent with an inner radius of $1.25$ mm and outer radius of $1.75$ mm with respect to the lumen center.
The detailed material parameters reported in \cref{tab:3-material} are taken from Gasser et al. \cite{gasser2007modeling}. Due to the symmetry, only half of the model is analyzed.

The only boundary conditions are the nodal displacements of the inner luminal nodes. A radial displacement loading is imposed on each node from its initial position, $r_i=1.0$ mm, to give a final deformed radius, $r_f=1.4$ mm, and to maintain the deformation in the healing process. 
The displacement loading is applied within 100 steps and G\&R is assumed to be started after displacement loading. 
The growth limit is set to the value of the determinant of the driven force of growth when the radius of lumen is $r_l=1.2$ mm, as $r_{gi}=\| (\bm q_{gi})\|_ {r_l=1.2\ \rm{mm}}$.

\begin{figure}[htbp]
	\centering
	\includegraphics[scale=0.65]{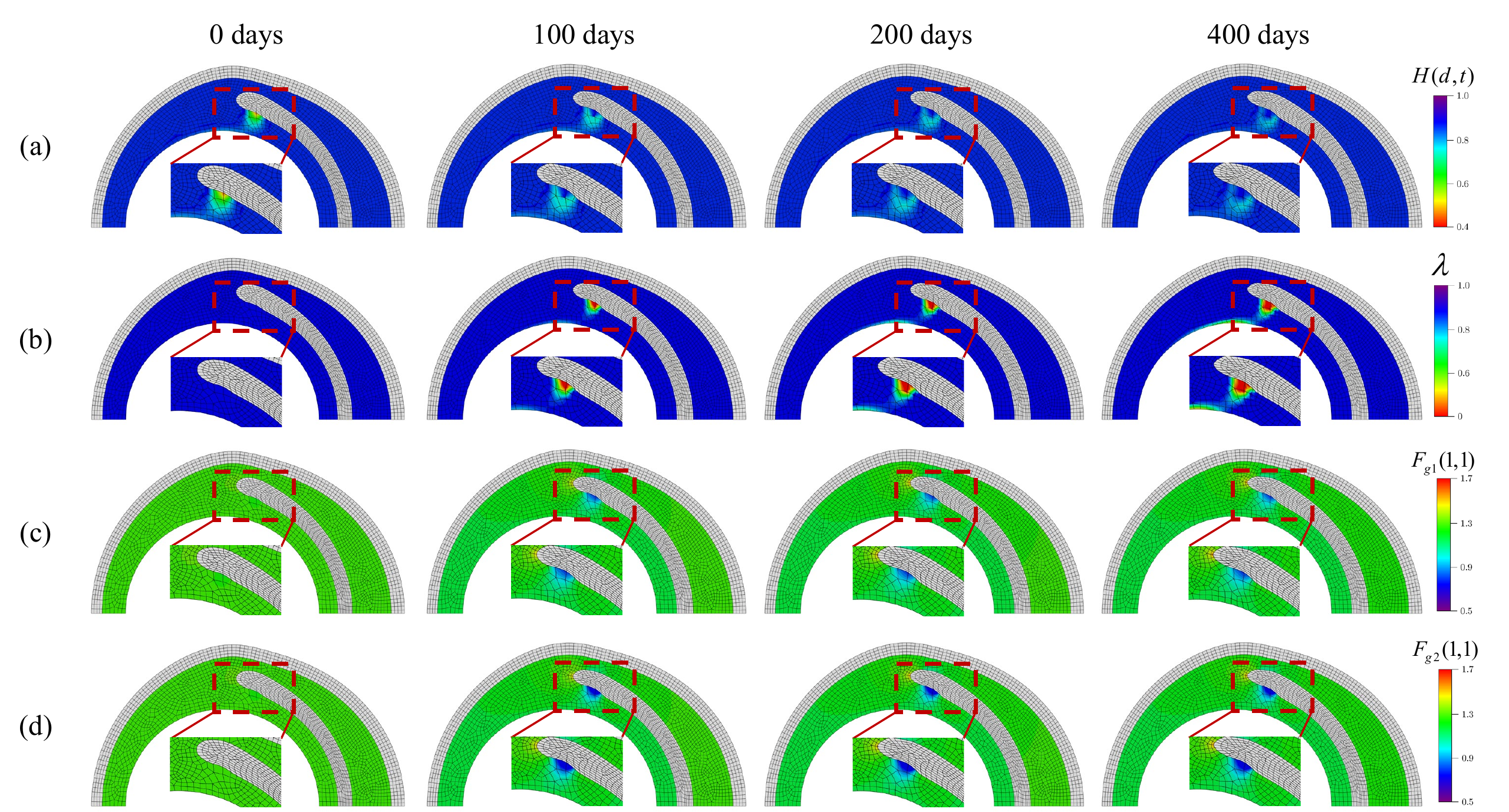}
	\caption{The contours for different parameters in healing process at different times. (a) The healing parameter $H(d,t)$, (b) the volume ratio of newly deposited part $\lambda$, (c) the component of the growth deformation for damaged part $F_{g1}(1,1)$, and (d) the component of the growth deformation for the newly deposited part $F_{g2}(1,1)$.}
	\label{fig:3-different-parameter-contour}
\end{figure}

The performance of the proposed model is tested by simulating the variations in four healing related parameters, $H(d,t)$, $\lambda$, $F_{g1}(1,1)$ and $F_{g2}(1,1)$, as shown in \cref{fig:3-different-parameter-contour}. The results of $H(d,t)$ in Fig. 14(a) show that the damage initially mainly occurs at the fiber cap in the shoulder of the plaque, the results agree with the review report of Holzapfel et al. \cite{2014Computational}.
The variation of $H(d,t)$ from 0 day to 400 day shown in Fig. 14(a) demonstrates that the damage in balloon angioplasty can be partly healed over a long-term time, for instance, the minimum value of the healing parameter $H(d,t)$ in the entire domain is increasing from 0.46 to 0.83 in this example. Moreover, it is interesting that our proposed model provides the variations of more parameters during the healing process at the same time, for instance, the results of $\lambda$ shown Fig. 14(b) indicates that the position where the new tissue is produced is almost the same with the position where the damage occurred. The component of the growth deformation for the original $F_{g1}(1,1)$ and the newly deposited tissues $F_{g2}(1,1)$ are also shown in Fig. 14(c) and Fig. 14(d), respectively, which illustrate the influence of growth on deformation in the healing process. 

\begin{figure}[htbp]
	\subfigure[]{
		\begin{minipage}{0.5\linewidth}
			\centering
			\includegraphics[scale=0.8]{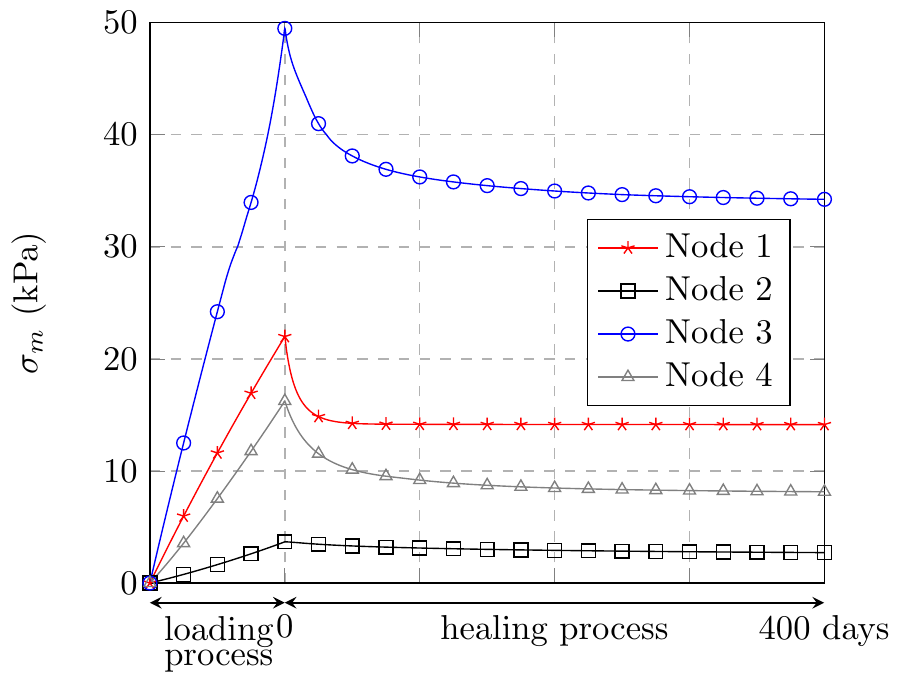}
			\label{fig:3-stress-different-node}
		\end{minipage}
	}
	\subfigure[]{
		\begin{minipage}{0.5\linewidth}
			\centering
			\includegraphics[scale=0.8]{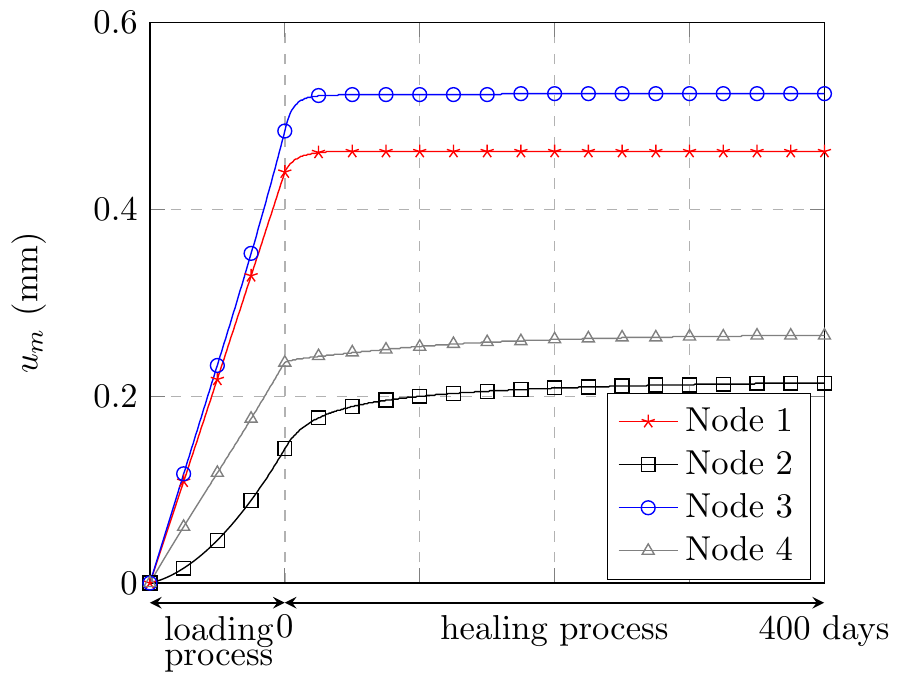}
			\label{fig:3-displacement-different-node}
		\end{minipage}
	}
	\caption{Influence of the inflation size at different locations in balloon angioplasty. (a) The variation of the Von Mises stress $\sigma_m$ with time, and (b) the variation of the displacement $u_m$ with time.}
	\label{fig:3-stress-displacement}
\end{figure}

To further investigate the evolution of G\&R at specific positions, four nodes (the locations shown in \cref{fig:ex.3}) are selected, and the Von Mises stress $\sigma_m$ and the magnitude of displacement $u_m=\sqrt{u_x^2+u_y^2}$ are calculated as shown in \cref{fig:3-stress-displacement}.
The curves of the variations of $\sigma_m$ and $u_m$ with time shown in \cref{fig:3-stress-displacement} illustrate that the proposed model works well and can converge to the homeostatic state.

\begin{landscape}
	\begin{figure}[htbp]
		\centering
		\includegraphics[scale=0.5]{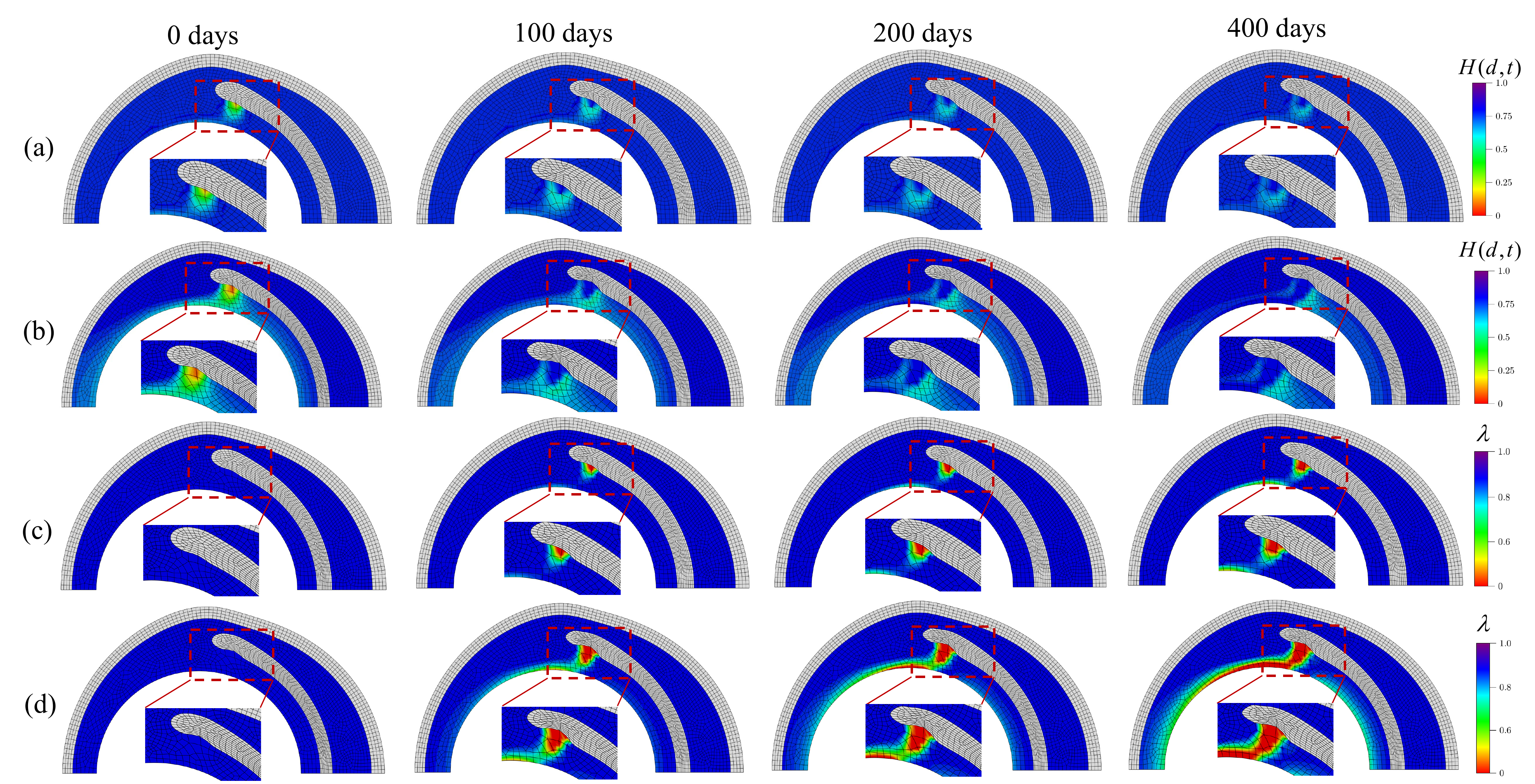}
		\caption{The contours for different parameters and inflation size in healing process at different times. (a) The healing parameter $H(d,t)$ when $r_f=1.40\ \rm mm$, (b) the healing parameter $H(d,t)$ when $r_f=1.48\ \rm mm$, (c) the newly deposited part $\lambda$ when $r_f=1.40\ \rm mm$, (b) the newly deposited part $\lambda$ when $r_f=1.48\ \rm mm$.}
		\label{fig:3-different-fd}
	\end{figure}
\end{landscape}

\begin{figure}[htbp]
	\centering
	\includegraphics[scale=1.0]{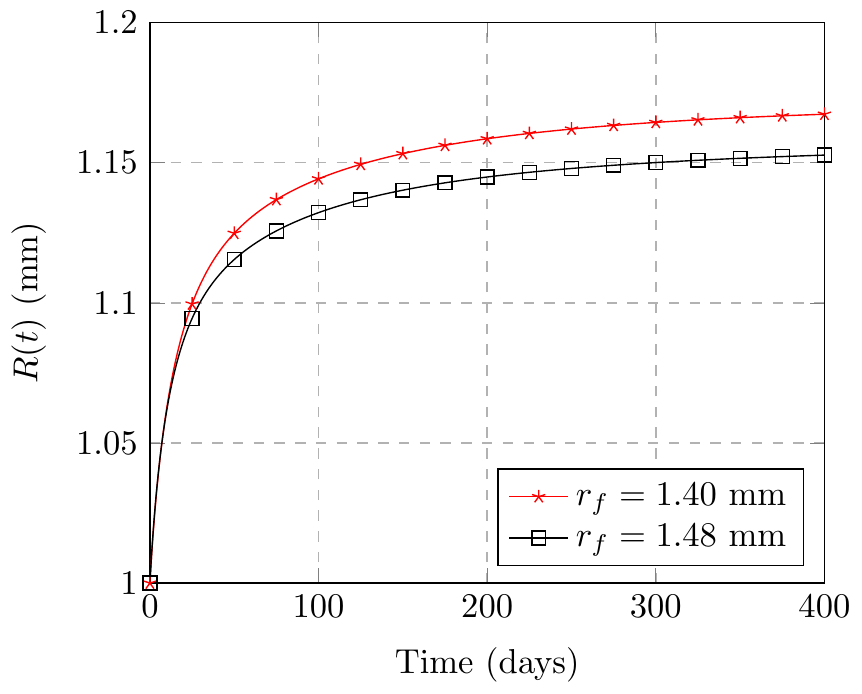}
	\caption{The variations of $R(t)$ for different inflation size in balloon angioplasty.}
	\label{fig:3-displacement-different-r}
\end{figure}

As the inflation size is the critical indicator in balloon angioplasty \cite{tenaglia1997intravascular}, two inflation sizes are tested to investigate the influence of the inflation size on G\&R, as the contours of the healing parameter $H(d,t)$ and the the volume ratio of the newly deposited part $\lambda$ are shown in \cref{fig:3-different-fd}.
The minimum value of the healing parameter $H(d,t)$ at $[0,100,200,400]$ days is $[0.46,0.82,0.83,0.83]$ when $r_f=1.40\ \rm mm$ compared with the corresponding results of $H(d,t)$ is $[0.17,0.74,0.77,0.79]$ when $r_f=1.48\ \rm mm$.
Although a faster healing speed can be observed at the beginning of the healing process when $r_f=1.48\ \rm mm$, a larger area of unrecoverable damage remains when healing is completed.
Therefore, the long-term evolution of the damage is also important and must be considered.
\cref{fig:3-displacement-different-r} shows the evolution of the normalized outer artery radius $R(t)$ in the healing process, obtained by calculating the time-dependent ratio of the outer radius of the artery $R(t)=\frac{r_{oa}(t)}{r_{oa}(t=0)}$, where $r_{oa}$ is the outer radius of the artery.
The variations of $R(t)$ shown in \cref{fig:3-displacement-different-r} demonstrates that the artery wall gradually changes to be thicker during the healing process and finally converges to a stable thickness at homeostatic state, this phenomenon is similar to the computational results by Braeu et al. in simulating of the thickening of arterial wall in hypertension \cite{braeu2017homogenized}.
It can be explained that thickening of the wall helps restore a homeostatic state, as it decreases the wall stress back to the initial level \cite{braeu2017homogenized}. 
A smaller deformation for $r_f=1.48\ \rm mm$ can be illustrated by a larger unrecoverable damage, which means that litter displacement is needed to recover to the homeostatic stress state, similar to the results in Example 2.

\section{Conclusions}
\label{S:4}
Based on the framework of thermodynamics, a new unified continuum damage model of the healing of soft biological tissues is proposed for the first time in this paper.
Different from the existing damage models of soft tissue healing, all the parameters related to the healing process can be regarded as the internal variables, in the same way as the damage variable. Therefore, the evolution of these healing parameters can be strictly derived based on the theory of thermodynamics, and thus, no \textit{ad hoc} equations are required as in the existing healing models. By virtue of the proposed unified damage models, the difficultly caused by the models available in the literature, for example, their disregard of continuum mechanical requirements such as that material frame indifference, explicit time dependence of material parameters, or unclear meaning of parameters, can be well overcome.

The proposed unified continuum damage model is validated by three representative numerical examples.
The basic performance of the proposed model is shown through a uniaxial tension scenario, the results of which show that the proposed model can well simulate the healing process, including the occurrence of damage and the recovery process. In addition, the evolution of the volume ratio of the newly deposited tissue and the growth deformation can be well illustrated. 
The nonlocal healing of the proposed model is achieved by a combination of the gradient terms,  good mesh independence is shown in the open-hole plate scenario, and the evolution of the volume ratios and growth deformations for both the original part and the newly deposited part for soft tissue is illustrated. 
The good potential of the method is demonstrated by a case of balloon angioplasty in the atherosclerotic artery with a fiber cap, where the long-term healing process in soft biological tissues after damage is simulated.
The numerical results of the proposed model agree well with the existing works in indicating the occurring position of damage for artery \cite{2014Computational} and predicting the trend of variation for arterial thickness in healing process \cite{braeu2017homogenized}.

The presented model is limited to 2D cases and to isotropic hyperelastic models. As collagen fibers are essential in the healing of soft tissue, the development of a 3D anisotropic model is currently in progress to address more realistic applications.
The identification of newly defined healing parameters in the proposed unified model is also critical for applications to practical problems and is currently underway. 

In summary, a new unified continuum damage model for the healing of soft biological tissues is presented in this paper. The evolution equations of healing parameters are derived based on the theory of thermodynamics. The proposed model provides a concise and rigorous framework for the establishment of a constitutive relationship and an \textit{in silico} simulation of the healing of soft biological tissues with newly derived parameters having clear physical interpretations. The proposed model will be useful in simulating the entire surgery and recovery process of individual patients based on CT or MRI data, particularly in evaluating the risks and probability of carrying out surgical intervention.

\section*{Declaration of Competing Interest}
The authors declare no competing interests.

\section*{CRediT author statement}
{\bf Di Zuo:} Software; Validation; Writing-Original Draft. 
{\bf Yiqian He:} Supervision; Conceptualization; Writing-Review \& Editing.
{\bf Stéphane Avril:} Writing-Review \& Editing.
{\bf Haitian Yang:} Writing-Review \& Editing.
{\bf Klaus Hackl:} Conceptualization; Methodology; Writing-Review \& Editing.

\section*{Acknowledgments}
The research leading to this paper was funded by the NSFC Grant [12072063], ERC-2014-CoG-BIOLOCHANICS
[647067], grants from the State Key Laboratory of Structural Analysis for Industrial Equipment [GZ19105, S18402],
the Liaoning Provincial Natural Science Foundation [2020-MS-110]. Klaus Hackl gratefully acknowledges financial support by Dalian University of Technology via a Haitian Scholarship.

\clearpage
\bibliographystyle{model1-num-names.bst}
\bibliography{reference.bib}

\begin{thebibliography}{25}
\expandafter\ifx\csname natexlab\endcsname\relax\def\natexlab#1{#1}\fi
\providecommand{\bibinfo}[2]{#2}
\ifx\xfnm\relax \def\xfnm[#1]{\unskip,\space#1}\fi
\bibitem[{Rao(2011)}]{rao2011modeling}
\bibinfo{author}{I.~Rao},
\newblock \bibinfo{title}{Modeling of growth and remodeling in soft biological
  tissues with multiple constituents},
\newblock \bibinfo{journal}{Mechanics Research Communications}
  \bibinfo{volume}{38} (\bibinfo{year}{2011}) \bibinfo{pages}{24--28}.
\bibitem[{Gasser(2017)}]{Gasser2017}
\bibinfo{author}{T.~C. Gasser}, \bibinfo{title}{Damage in Vascular Tissues and
  Its Modeling}, \bibinfo{publisher}{Springer International Publishing},
  \bibinfo{address}{Cham}, pp. \bibinfo{pages}{85--118}.
\bibitem[{Valero et~al.(2015)Valero, Javierre, Garc{\'\i}a-Aznar,
  G{\'o}mez-Benito, and Menzel}]{valero2015modeling}
\bibinfo{author}{C.~Valero}, \bibinfo{author}{E.~Javierre},
  \bibinfo{author}{J.~Garc{\'\i}a-Aznar},
  \bibinfo{author}{M.~G{\'o}mez-Benito}, \bibinfo{author}{A.~Menzel},
\newblock \bibinfo{title}{Modeling of anisotropic wound healing},
\newblock \bibinfo{journal}{Journal of the Mechanics and Physics of Solids}
  \bibinfo{volume}{79} (\bibinfo{year}{2015}) \bibinfo{pages}{80--91}.
\bibitem[{Comellas et~al.(2016)Comellas, Gasser, Bellomo, and
  Oller}]{comellas2016homeostatic}
\bibinfo{author}{E.~Comellas}, \bibinfo{author}{T.~C. Gasser},
  \bibinfo{author}{F.~J. Bellomo}, \bibinfo{author}{S.~Oller},
\newblock \bibinfo{title}{A homeostatic-driven turnover remodelling
  constitutive model for healing in soft tissues},
\newblock \bibinfo{journal}{Journal of the Royal Society Interface}
  \bibinfo{volume}{13} (\bibinfo{year}{2016}) \bibinfo{pages}{20151081}.
\bibitem[{Cumming et~al.(2009)Cumming, McElwain, and
  Upton}]{cumming2009mathematical}
\bibinfo{author}{B.~D. Cumming}, \bibinfo{author}{D.~McElwain},
  \bibinfo{author}{Z.~Upton},
\newblock \bibinfo{title}{A mathematical model of wound healing and subsequent
  scarring},
\newblock \bibinfo{journal}{Journal of The Royal Society Interface}
  \bibinfo{volume}{7} (\bibinfo{year}{2009}) \bibinfo{pages}{19--34}.
\bibitem[{Buganza~Tepole and Kuhl(2016)}]{buganza2016computational}
\bibinfo{author}{A.~Buganza~Tepole}, \bibinfo{author}{E.~Kuhl},
\newblock \bibinfo{title}{Computational modeling of chemo-bio-mechanical
  coupling: a systems-biology approach toward wound healing},
\newblock \bibinfo{journal}{Computer methods in biomechanics and biomedical
  engineering} \bibinfo{volume}{19} (\bibinfo{year}{2016})
  \bibinfo{pages}{13--30}.
\bibitem[{Javierre et~al.(2009)Javierre, Moreo, Doblar{\'e}, and
  Garc{\'\i}a-Aznar}]{javierre2009numerical}
\bibinfo{author}{E.~Javierre}, \bibinfo{author}{P.~Moreo},
  \bibinfo{author}{M.~Doblar{\'e}}, \bibinfo{author}{J.~Garc{\'\i}a-Aznar},
\newblock \bibinfo{title}{Numerical modeling of a mechano-chemical theory for
  wound contraction analysis},
\newblock \bibinfo{journal}{International journal of solids and structures}
  \bibinfo{volume}{46} (\bibinfo{year}{2009}) \bibinfo{pages}{3597--3606}.
\bibitem[{Schugart et~al.(2008)Schugart, Friedman, Zhao, and
  Sen}]{schugart2008wound}
\bibinfo{author}{R.~C. Schugart}, \bibinfo{author}{A.~Friedman},
  \bibinfo{author}{R.~Zhao}, \bibinfo{author}{C.~K. Sen},
\newblock \bibinfo{title}{Wound angiogenesis as a function of tissue oxygen
  tension: a mathematical model},
\newblock \bibinfo{journal}{Proceedings of the National Academy of Sciences}
  \bibinfo{volume}{105} (\bibinfo{year}{2008}) \bibinfo{pages}{2628--2633}.
\bibitem[{Kuhl et~al.(2005)Kuhl, Garikipati, Arruda, and
  Grosh}]{kuhl2005remodeling}
\bibinfo{author}{E.~Kuhl}, \bibinfo{author}{K.~Garikipati},
  \bibinfo{author}{E.~M. Arruda}, \bibinfo{author}{K.~Grosh},
\newblock \bibinfo{title}{Remodeling of biological tissue: mechanically induced
  reorientation of a transversely isotropic chain network},
\newblock \bibinfo{journal}{Journal of the Mechanics and Physics of Solids}
  \bibinfo{volume}{53} (\bibinfo{year}{2005}) \bibinfo{pages}{1552--1573}.
\bibitem[{Humphrey and Rajagopal(2002)}]{humphrey2002constrained}
\bibinfo{author}{J.~Humphrey}, \bibinfo{author}{K.~Rajagopal},
\newblock \bibinfo{title}{A constrained mixture model for growth and remodeling
  of soft tissues},
\newblock \bibinfo{journal}{Mathematical models and methods in applied
  sciences} \bibinfo{volume}{12} (\bibinfo{year}{2002})
  \bibinfo{pages}{407--430}.
\bibitem[{Latorre and Humphrey(2018)}]{latorre2018critical}
\bibinfo{author}{M.~Latorre}, \bibinfo{author}{J.~D. Humphrey},
\newblock \bibinfo{title}{Critical roles of time-scales in soft tissue growth
  and remodeling},
\newblock \bibinfo{journal}{APL bioengineering} \bibinfo{volume}{2}
  (\bibinfo{year}{2018}) \bibinfo{pages}{026108}.
\bibitem[{Dimitrijevic and Hackl(2008)}]{dimitrijevic2008method}
\bibinfo{author}{B.~Dimitrijevic}, \bibinfo{author}{K.~Hackl},
\newblock \bibinfo{title}{A method for gradient enhancement of continuum damage
  models},
\newblock \bibinfo{journal}{Technische Mechanik} \bibinfo{volume}{28}
  (\bibinfo{year}{2008}) \bibinfo{pages}{43--52}.
\bibitem[{He et~al.(2019)He, Zuo, Hackl, Yang, Mousavi, and
  Avril}]{he2019gradient}
\bibinfo{author}{Y.~He}, \bibinfo{author}{D.~Zuo}, \bibinfo{author}{K.~Hackl},
  \bibinfo{author}{H.~Yang}, \bibinfo{author}{S.~J. Mousavi},
  \bibinfo{author}{S.~Avril},
\newblock \bibinfo{title}{Gradient-enhanced continuum models of healing in
  damaged soft tissues},
\newblock \bibinfo{journal}{Biomechanics and modeling in mechanobiology}
  (\bibinfo{year}{2019}) \bibinfo{pages}{1--18}.
\bibitem[{Zuo et~al.(2020)Zuo, Avril, Yang, Mousavi, Hackl, and
  He}]{zuo2020threedimensional}
\bibinfo{author}{D.~Zuo}, \bibinfo{author}{S.~Avril},
  \bibinfo{author}{H.~Yang}, \bibinfo{author}{S.~J. Mousavi},
  \bibinfo{author}{K.~Hackl}, \bibinfo{author}{Y.~He},
\newblock \bibinfo{title}{Three-dimensional numerical simulation of soft-tissue
  wound healing using constrained-mixture anisotropic hyperelasticity and
  gradient-enhanced damage mechanics},
\newblock \bibinfo{journal}{Journal of the Royal Society Interface}
  \bibinfo{volume}{17} (\bibinfo{year}{2020}) \bibinfo{pages}{20190708}.
\bibitem[{Frank et~al.(1999{\natexlab{a}})Frank, Shrive, Hiraoka, Nakamura,
  Kaneda, and Hart}]{frank1999optimisation}
\bibinfo{author}{C.~B. Frank}, \bibinfo{author}{N.~G. Shrive},
  \bibinfo{author}{H.~Hiraoka}, \bibinfo{author}{N.~Nakamura},
  \bibinfo{author}{Y.~Kaneda}, \bibinfo{author}{D.~A. Hart},
\newblock \bibinfo{title}{Optimisation of the biology of soft tissue repair},
\newblock \bibinfo{journal}{Journal of Science and Medicine in Sport}
  \bibinfo{volume}{2} (\bibinfo{year}{1999}{\natexlab{a}})
  \bibinfo{pages}{190--210}.
\bibitem[{Frank et~al.(1999{\natexlab{b}})Frank, Hart, and
  Shrive}]{frank1999molecular}
\bibinfo{author}{C.~B. Frank}, \bibinfo{author}{D.~A. Hart},
  \bibinfo{author}{N.~G. Shrive},
\newblock \bibinfo{title}{Molecular biology and biomechanics of normal and
  healing ligaments—a review},
\newblock \bibinfo{journal}{Osteoarthritis and Cartilage} \bibinfo{volume}{7}
  (\bibinfo{year}{1999}{\natexlab{b}}) \bibinfo{pages}{130--140}.
\bibitem[{Dimitrijevic and Hackl(2011)}]{dimitrijevic2011regularization}
\bibinfo{author}{B.~Dimitrijevic}, \bibinfo{author}{K.~Hackl},
\newblock \bibinfo{title}{A regularization framework for damage--plasticity
  models via gradient enhancement of the free energy},
\newblock \bibinfo{journal}{International Journal for Numerical Methods in
  Biomedical Engineering} \bibinfo{volume}{27} (\bibinfo{year}{2011})
  \bibinfo{pages}{1199--1210}.
\bibitem[{Nolan et~al.(2014)Nolan, Gower, Destrade, Ogden, and
  McGarry}]{nolan2014robust}
\bibinfo{author}{D.~Nolan}, \bibinfo{author}{A.~Gower},
  \bibinfo{author}{M.~Destrade}, \bibinfo{author}{R.~Ogden},
  \bibinfo{author}{J.~McGarry},
\newblock \bibinfo{title}{A robust anisotropic hyperelastic formulation for the
  modelling of soft tissue},
\newblock \bibinfo{journal}{Journal of the mechanical behavior of biomedical
  materials} \bibinfo{volume}{39} (\bibinfo{year}{2014})
  \bibinfo{pages}{48--60}.
\bibitem[{Waffenschmidt et~al.(2014)Waffenschmidt, Polindara, Menzel, and
  Blanco}]{waffenschmidt2014gradient}
\bibinfo{author}{T.~Waffenschmidt}, \bibinfo{author}{C.~Polindara},
  \bibinfo{author}{A.~Menzel}, \bibinfo{author}{S.~Blanco},
\newblock \bibinfo{title}{A gradient-enhanced large-deformation continuum
  damage model for fibre-reinforced materials},
\newblock \bibinfo{journal}{Computer Methods in Applied Mechanics and
  Engineering} \bibinfo{volume}{268} (\bibinfo{year}{2014})
  \bibinfo{pages}{801--842}.
\bibitem[{Polindara et~al.(2017)Polindara, Waffenschmidt, and
  Menzel}]{polindara2017computational}
\bibinfo{author}{C.~Polindara}, \bibinfo{author}{T.~Waffenschmidt},
  \bibinfo{author}{A.~Menzel},
\newblock \bibinfo{title}{A computational framework for modelling
  damage-induced softening in fibre-reinforced materials--application to
  balloon angioplasty},
\newblock \bibinfo{journal}{International Journal of Solids and Structures}
  \bibinfo{volume}{118} (\bibinfo{year}{2017}) \bibinfo{pages}{235--256}.
\bibitem[{Loree et~al.(1992)Loree, Kamm, Stringfellow, and
  Lee}]{loree1992effects}
\bibinfo{author}{H.~M. Loree}, \bibinfo{author}{R.~Kamm},
  \bibinfo{author}{R.~Stringfellow}, \bibinfo{author}{R.~T. Lee},
\newblock \bibinfo{title}{Effects of fibrous cap thickness on peak
  circumferential stress in model atherosclerotic vessels.},
\newblock \bibinfo{journal}{Circulation research} \bibinfo{volume}{71}
  (\bibinfo{year}{1992}) \bibinfo{pages}{850--858}.
\bibitem[{Gasser and Holzapfel(2007)}]{gasser2007modeling}
\bibinfo{author}{T.~C. Gasser}, \bibinfo{author}{G.~A. Holzapfel},
\newblock \bibinfo{title}{Modeling plaque fissuring and dissection during
  balloon angioplasty intervention},
\newblock \bibinfo{journal}{Annals of biomedical engineering}
  \bibinfo{volume}{35} (\bibinfo{year}{2007}) \bibinfo{pages}{711--723}.
\bibitem[{Holzapfel et~al.(2014)Holzapfel, Mulvihill, Cunnane, and
  Walsh}]{2014Computational}
\bibinfo{author}{G.~A. Holzapfel}, \bibinfo{author}{J.~J. Mulvihill},
  \bibinfo{author}{E.~M. Cunnane}, \bibinfo{author}{M.~T. Walsh},
\newblock \bibinfo{title}{Computational approaches for analyzing the mechanics
  of atherosclerotic plaques: A review},
\newblock \bibinfo{journal}{Journal of Biomechanics} \bibinfo{volume}{47}
  (\bibinfo{year}{2014}) \bibinfo{pages}{859--869}.
\bibitem[{Tenaglia et~al.(1997)}]{tenaglia1997intravascular}
\bibinfo{author}{A.~N. Tenaglia}, et~al.,
\newblock \bibinfo{title}{Intravascular ultrasound and balloon percutaneous
  transluminal coronary angioplasty},
\newblock \bibinfo{journal}{Cardiology clinics} \bibinfo{volume}{15}
  (\bibinfo{year}{1997}) \bibinfo{pages}{31--38}.
\bibitem[{Braeu et~al.(2017)Braeu, Seitz, Aydin, and
  Cyron}]{braeu2017homogenized}
\bibinfo{author}{F.~Braeu}, \bibinfo{author}{A.~Seitz},
  \bibinfo{author}{R.~Aydin}, \bibinfo{author}{C.~Cyron},
\newblock \bibinfo{title}{Homogenized constrained mixture models for
  anisotropic volumetric growth and remodeling},
\newblock \bibinfo{journal}{Biomechanics and modeling in mechanobiology}
  \bibinfo{volume}{16} (\bibinfo{year}{2017}) \bibinfo{pages}{889--906}.

\end{thebibliography}
	
\clearpage
\listoftables

\clearpage
\begin{table}[htbp]
	\centering
	\caption{Material parameters for the uniaxial tension. }
	\resizebox{\textwidth}{!}{
		\begin{tabular}{lllll}
			\toprule
			Type  & Description & Symbol & Values & Units \\
			\midrule
			\multirow{4}[1]{*}{Hyperelastic} & \multirow{2}[1]{*}{Shear modulus} & $\mu_1$ & 1     & MPa \\
			&       & $\mu_2$ & 1     & MPa \\
			& \multirow{2}[0]{*}{Bulk modulus} & $k_1$  & 1     & MPa \\
			&       & $k_2$  & 1     & MPa \\
			\midrule
			\multirow{4}{*}{Damage} 
			& Damage threshold 			&   $\kappa_d$    	& 0.2	 	& MPa \\       
			& Degree of regularisation 	&    $c_d$    		& 1     	& MPa$\cdot \rm mm^{2}$ \\     
			& Penalty parameter 		&   $\beta_d$    	& 0.001 	& MPa  \\     
			& (Non)local switch 		&    $\gamma_d$  	& 1     	& - \\
			\hline\noalign{\smallskip}
			\multirow{2}[0]{*}{Remodeling} & Remodeling rate & $M_{rm}$ & [0.01,0.02,0.05]  & $day^{-1}$ \\
			& Irreversible stiffness loss & $\eta$ & [0,0.2,0.5,1.0]     & - \\
			\midrule
			\multirow{4}[1]{*}{Growth} & \multirow{2}[0]{*}{Growth rate} & $M_{g1}$   & [0.01,0.02,0.05] & $day^{-1}$ \\
			&       & $M_{g2}$   & [0.01,0.02,0.05] & $day^{-1}$ \\
			& \multirow{2}[1]{*}{Growth limit} & $r_{g1}$   & -     & - \\
			&       & $r_{g2}$   & -     & - \\
			\bottomrule
		\end{tabular}%
	}
	\label{tab:1-material}%
\end{table}%

\clearpage
\begin{table}[htbp]
	\centering
	\caption{Material parameters for the open-hole plate. }
	\begin{tabular}{lllll}
		\toprule
		Type  & Description & Symbol & Values & Units \\
		\midrule
		\multirow{3}{*}{Geometry}&
		Height		 					& $H$ 			& 100		 	& mm				\\
		&
		Width  					    & $W$ 	     	& 100			& mm				\\
		&
		Diameter  					    & $R$ 	     	& 50			& mm				\\
		\hline\noalign{\smallskip}
		\multirow{4}[1]{*}{Hyperelastic} & \multirow{2}[1]{*}{Shear modulus} & $\mu_1$ & 1     & MPa \\
		&       & $\mu_2$ & 1     & MPa \\
		& \multirow{2}[0]{*}{Bulk modulus} & $k_1$  & 40     & MPa \\
		&       & $k_2$  & 40     & MPa \\
		\midrule
		\multirow{4}{*}{Damage} 
		& Damage threshold 			&   $\kappa_d$    	& 0.01	 	& MPa \\       
		& Degree of regularization 	&    $c_d$    		& 1     	& MPa$\cdot \rm mm^{2}$ \\     
		& Penalty parameter 		&   $\beta_d$    	& 1 	& MPa  \\     
		& (Non)local switch 		&    $\gamma_d$  	& 1     	& - \\
		\hline\noalign{\smallskip}
		\multirow{2}[0]{*}{Remodeling} & Remodeling rate & $M_{rm}$ & [0.1,1.0]  & $day^{-1}$ \\
		& Irreversible stiffness loss & $\eta$ & 0     & - \\
		\midrule
		\multirow{4}[1]{*}{Growth} & \multirow{2}[0]{*}{Growth rate} & $M_{g1}$   &[0.01,0.03] & $day^{-1}$ \\
		&       & $M_{g2}$   & [0.01,0.03] & $day^{-1}$ \\
		& \multirow{2}[1]{*}{Growth limit} & $r_{g1}$   & -     & - \\
		&       & $r_{g2}$   & -     & - \\
		\bottomrule
	\end{tabular}%
	\label{tab:2-material}%
\end{table}%

\clearpage
\begin{table}[htbp]
	\centering
	\caption{Material parameters for the balloon angioplasty. }
	\begin{tabular}{lllll}
		\toprule
		Type  & Description & Symbol & Values & Units \\
		\midrule
		\multirow{8}[1]{*}{Hyperelastic} & \multirow{4}[1]{*}{Shear modulus} & $\mu_a$ & 15     & kPa \\
		&       & $\mu_{p1}$ & 78.9     & kPa \\
		&       & $\mu_{p2}$ & 78.9     & kPa \\
		&       & $\mu_l$ & 0.1     & kPa \\
		& \multirow{4}[0]{*}{Bulk modulus} & $k_a$  & 4     & kPa \\
		&       & $k_{p1}$  & 23.7     & kPa \\
		&       & $k_{p2}$ & 23.7     & kPa \\
		&       & $k_l$ & 0.5     & kPa \\
		\midrule
		\multirow{4}{*}{Damage} 
		& Damage threshold 			&   $\kappa_d$    	& 5	 	& kPa \\       
		& Degree of regularization 	&    $c_d$    		& 1    	& kPa$\cdot \rm mm^{2}$ \\     
		& Penalty parameter 		&   $\beta_d$    	& 20 	& kPa  \\     
		& (Non)local switch 		&    $\gamma_d$  	& 1     	& - \\
		\hline\noalign{\smallskip}
		\multirow{2}[0]{*}{Remodeling} & Remodeling rate & $M_{rm}$ &  0.1 & $day^{-1}$ \\
		& Irreversible stiffness loss & $\eta$ &    1.0  & - \\
		\midrule
		\multirow{4}[1]{*}{Growth} & \multirow{2}[0]{*}{Growth rate} & $M_{g1}$   &0.01  & $day^{-1}$ \\
		&      & $M_{g2}$   &   0.01& $day^{-1}$ \\
		& \multirow{2}[1]{*}{Growth limit} & $r_{g1}$   & -    & - \\
		&     & $r_{g2}$   & -     & - \\
		\bottomrule
	\end{tabular}%
	\label{tab:3-material}%
\end{table}%

\clearpage
\listoffigures

\clearpage

\clearpage

\clearpage

\clearpage

\clearpage

\clearpage

\clearpage

\clearpage

\clearpage

\clearpage

\clearpage

\clearpage

\clearpage
\clearpage

\clearpage

\clearpage

\clearpage

\end{document}